\documentclass{emulateapj}
\usepackage{graphicx}
\usepackage[usenames,dvipsnames]{color}

\shorttitle{Optical/NIR Photo-polarimetric study of RCW41}
\shortauthors{Santos, Roman-Lopes \& Franco}

\begin{document}

\title{A young stellar cluster within the RCW41 H{\sc ii} region: deep NIR photometry and Optical/NIR polarimetry{\Large $^\star$}}

\thanks{{\large $\star$} Based on observations collected at the European Southern Observatory (La Silla, Chile), 
 National Optical Astronomy Observatory (CTIO, Chile) and Observat\'orio
 do Pico dos Dias, operated by Laborat\'orio Nacional de Astrof\'\i sica 
 (LNA/MCT, Brazil).}

%
%
%

\author{F\'abio P. Santos\altaffilmark{1}, Alexandre Roman-Lopes\altaffilmark{2} \& Gabriel A. P. Franco\altaffilmark{1}}
\affil{\altaffilmark{1}Departamento de F\'isica--ICEx--UFMG, Caixa Postal 702, 30.123-970 \\
Belo Horizote, MG, Brazil; [fabiops;franco]@fisica.ufmg.br}
\affil{\altaffilmark{2}Departamento de Fisica -- Universidad de La Serena, Cisternas 1200, La Serena, Chile; roman@dfuls.cl}

\begin{abstract}

The RCW41 star-forming region is embedded within the Vela Molecular Ridge, hosting a 
massive stellar cluster surrounded by a conspicuous H{\sc ii} region. Understanding the role 
of interstellar magnetic fields and studying the newborn stellar population is crucial to 
build a consistent picture of the physical processes acting on this kind of environment.
We have carried out a detailed study of the interstellar polarization toward RCW41,
with data from an optical and near-infrared polarimetric survey.
Additionally, deep near-infrared images from the NTT 3.5\,m telescope have been used to 
study the photometric properties of the embedded young stellar cluster, revealing 
several YSO's candidates. By using a set of pre-main sequence isochrones, 
a mean cluster age in the range 2.5 - 5.0 million years was determined, 
and evidence of sequential star formation were revealed. 
An abrupt decrease in R-band polarization degree is noticed toward the central ionized area, 
probably due to low grain alignment efficiency 
caused by the turbulent environment and/or weak intensity of magnetic fields. The distortion 
of magnetic field lines exhibit a dual behavior, with the mean orientation outside the area 
approximately following the borders of the star-forming region, and directed radially toward 
the cluster inside the ionized area, in agreement with simulations of expanding H{\sc ii} regions.
The spectral dependence of polarization allowed a meaningful determination of the 
total-to-selective extinction ratio by fittings of the Serkowski relation. Furthermore, a large 
rotation of polarization angle as a function of wavelength is detected toward several 
embedded stars.

\end{abstract}


\section{Introduction}
\label{introduction}

   \begin{figure*}[t]
   \centering
   \includegraphics[width=\textwidth]{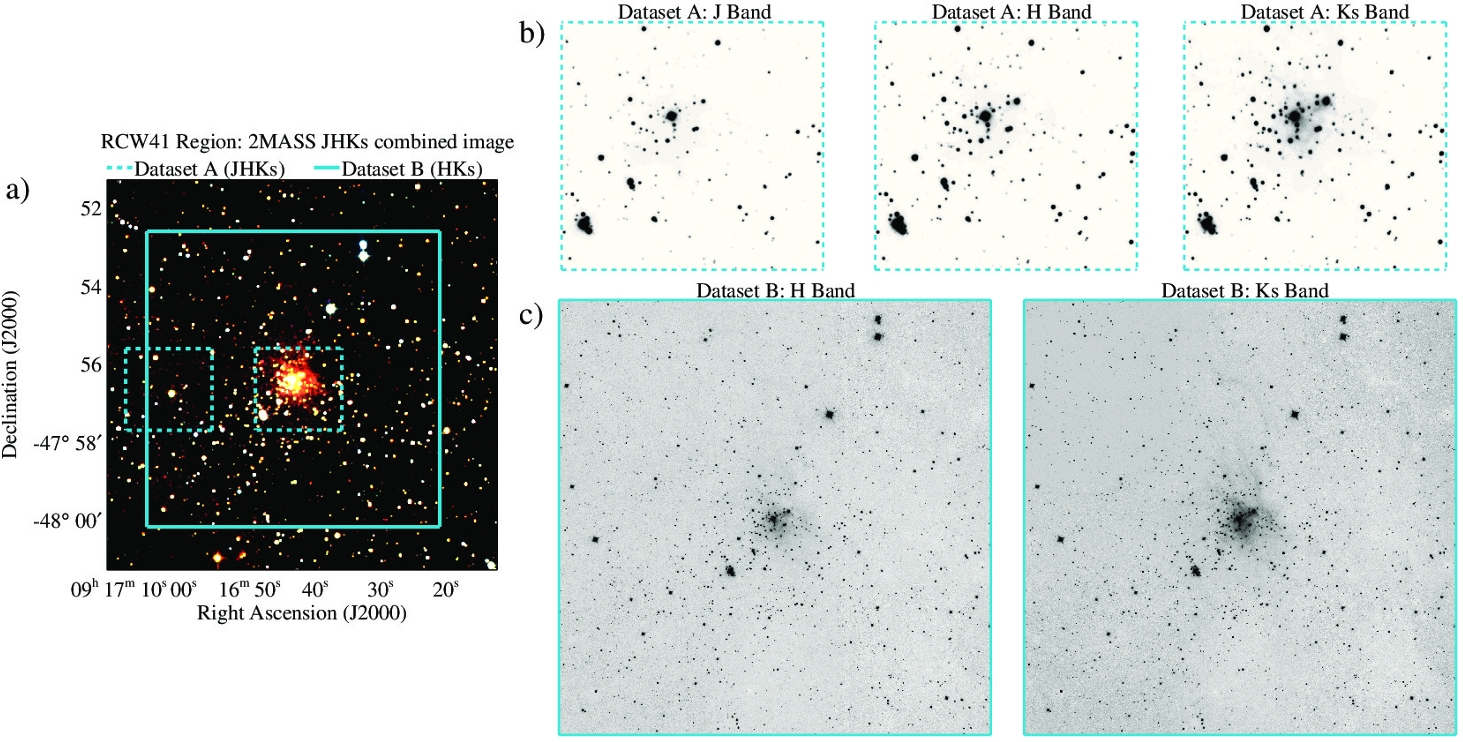}
      \caption{{\it (a)} Composite JHKs image (RGB) from the 2MASS survey showing
               the cluster in the direction of the IRAS 09149-4743 source,
               as well as the regions associated to our NIR photometry datasets:
               the dashed-line cyan rectangles indicates the regions related to Dataset A (J, H and Ks filters)
               and the solid-line cyan rectangle is related to Dataset B (H and Ks filters).
               {\it (b)} Images of the cluster region from Dataset A, at filters
               J ({\it left}), H ({\it middle}) and Ks ({\it right}).
               {\it (c)} Mosaics of the entire region of RCW41 covered by Dataset B at
               H ({\it left}) and Ks ({\it right}) filters.
              }
         \label{phot_datasets}
   \end{figure*}

   \begin{figure}[t]
   \centering
   \includegraphics[width=0.48\textwidth]{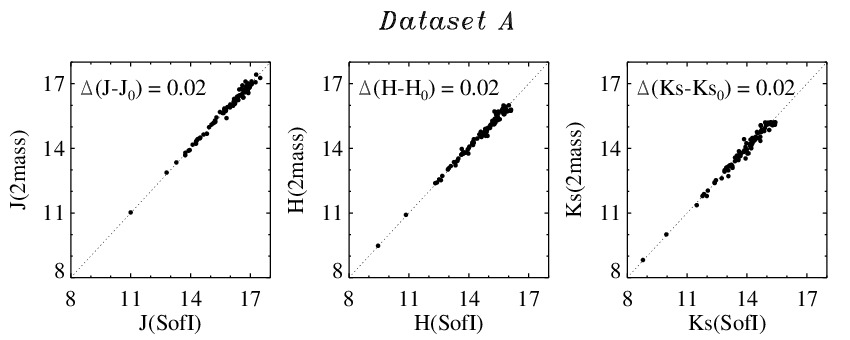} \\
   \includegraphics[width=0.48\textwidth]{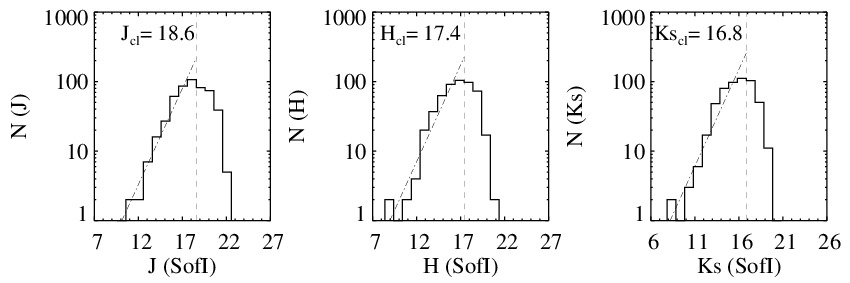} \\
   \includegraphics[width=0.32\textwidth]{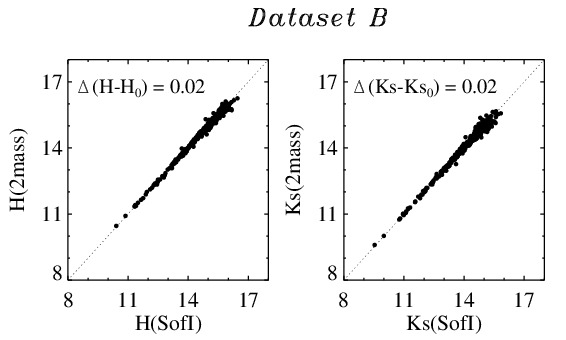} \\
   \includegraphics[width=0.32\textwidth]{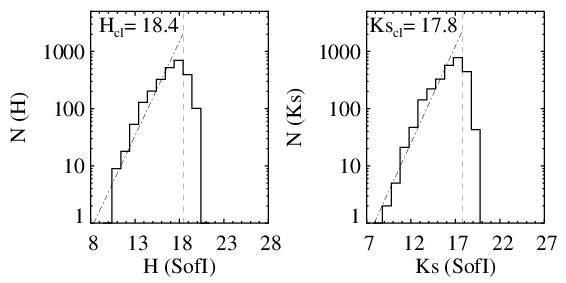} 
      \caption{
      Comparison between SofI and 2MASS magnitudes as well as magnitude histograms used to
      derive the completeness limits, shown {\it above} for Dataset A (J, H and Ks filters)
      and {\it below} for Dataset B (H and Ks filters).
      The comparison diagrams were used to determine the zero point corrections
      for each filter, and the respective zero point uncertainties
      are indicated inside each diagram. The dashed lines represents
      the ideal correspondence between the two systems, and were inserted in order
      to clarify the comparison. Only isolated and good quality photometry stars (flag ``A") from the 
      2MASS catalog were used. The magnitude histograms were used to calculate the 
      completeness limits (cl) to each filter from Datasets A and B, and are indicated by the vertical 
      dashed lines.
              }
         \label{comparison_completeness}
   \end{figure}
%

   \begin{figure*}[t]
   \centering
   \includegraphics[width=\textwidth]{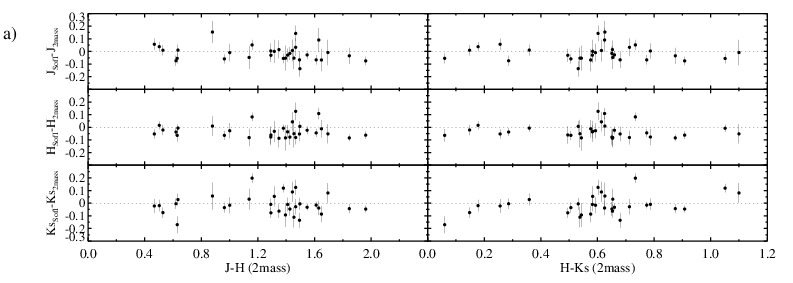} 
   \includegraphics[width=\textwidth]{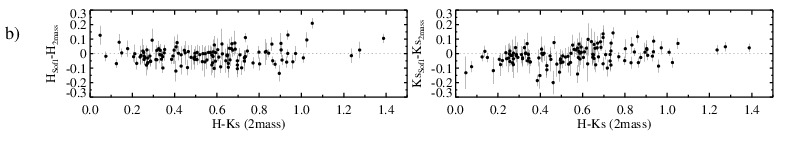}
      \caption{Diagrams related to Datasets A (a) and B (b), used to verify a possible color dependence
      associated to the photometric calibration of the SofI magnitudes based on the
      2MASS magnitudes. The horizontal dashed lines were inserted for clarity.
      The dispersion shows that no presence of color dependences are evident within our instrumental 
      uncertainties.
              }
         \label{colordep}
   \end{figure*}

It is well known that sites of massive star formation are mainly found inside heavily 
obscured dusty molecular cores, which are commonly associated to giant molecular cloud 
complexes \citep[e.g.,][]{lada2003,lada2010}. Photometric studies at near-infrared (NIR) 
wavelengths make possible to probe deep inside where star formation is taking place, with 
the very young stars frequently showing large infrared excesses due to the presence of 
warm dust from the associated circumstellar disks.

On the other hand, the study of the interstellar dust grain's interaction with the local magnetic 
fields, is a powerful tool to understand some of the physical processes acting in such 
environments. In this context, magnetic fields might play a key role in star formation. In fact, 
several physical processes, like the initial cloud collapse by the fragmentation of early stage 
star-forming cores, the channeling of interstellar material via ambipolar diffusion, the
disk and bipolar outflow formation, the angular momentum transport, etc, possibly are 
strongly affected by the strength and configuration of the local magnetic field components 
\citep{mestel1956,nakano1979,mou1981,shu1987,lizano1989,heitsch2004,girart2009}.
However, it is still unknown whether magnetic fields or interstellar turbulence provide the 
main supporting source against the cloud collapse, ultimately defining the star formation
efficiency \citep{padoan2004,crutcher2005,heiles2005,mckee2007}.

The formation of massive stars causes a large impact on the surrounding environment, 
ionizing the interstellar gas and providing a source of turbulence due to strong stellar winds 
and outflows from the newborn stellar population. Only recently, theoretical studies have 
attempted to describe a realistic view of the impact of the expansion of H{\sc ii} regions on the 
underlying structures of the magnetic field lines 
\citep{krumholz2007,peters2010,peters2011,arthur2011}.
Unfortunately, mappings of the magnetic field structures along the majority of Galactic H{\sc ii} 
regions are scarcely available, although these could serve as important tests and provide 
additional constraints to such models of star formation. 

An important method to map the sky-projected lines of magnetic field using optical or NIR
spectral bands, is from interstellar polarization by aligned dust grains, that serves as a
well known tracer. A detailed study of linear polarization toward H{\sc ii} regions and young stellar 
clusters can provide useful information regarding the structure of the Galactic magnetic 
field in these regions.

The Vela Molecular Ridge (VMR) is an interesting Galactic complex where star formation
is taking place over a wide range of masses \citep{liseau1992,lorenzetti1993,massi1999,massi2000,massi2003}.
It is located at the Galactic Plane roughly towards $l\sim265^{\circ}$, where one can see the 
presence of several large interstellar features, such as the Vela supernova remnant, the Gum Nebula, 
and other structures.
Beyond this ``local" structure, at the $1-2$\,kpc range, several young star clusters are found, 
as well as numerous optical H{\sc ii} regions with several signposts of embedded star formation such as H$_{2}$O masers
\citep{braz1983,zin1995}.

Among these structures, a giant molecular complex extends over $20^{\circ}$ in the southern sky. It can be 
identified by its strong CO emission, and was first studied by \citet{murphy1991}, who divided the main structure into four
regions, named $A$ to $D$.
Clouds towards these areas are located along a wide range of distances, from $700$ to $\sim 2000$ pc \citep{liseau1992}.
According to \citet{murphy1991} regions $A$, $C$ and $D$ are located quite closer to us ($\sim1$kpc), with the cloud $B$ 
being slightly more distant ($\sim2$kpc). Recently, a submillimetric survey carried out by \citet{olmi2009} detected 
approximately 140 proto- and pre-stellar cores in the direction of cloud $D$.

The IRAS 09149-4743 source is an example of young stellar cluster in this region. It is part 
of the cloud $A$, being located at an estimated distance of $1.3\pm0.2$\,kpc \citep{roman2009}. It shows 
several candidate young stellar objects (YSOs), as well as, two massive stars of spectral types 
O9{\sc v} and B0{\sc v}, as revealed by previous photometric and spectroscopic infrared 
surveys \citep{ortiz2007, roman2009}. 
Several radio and infrared observations are available in the direction 
of IRAS 09149-4743, tracing a number of chemical elements in this region, 
which are mainly related to the existence of the surrounding H{\sc ii} region 
and the massive star forming area. 
Ortiz et al. (2007) compiled a list of all observational data collected from the 
literature up to that date.
Furthermore, \citet{pirogov2007} reported the detection of CS, N$_{2}$H$^{+}$ and 1.2 mm dust 
continuum emissions toward IRAS 09149-4743. Specifically, the dust emission represent an almost 
spherical core approximately superposed to the stellar cluster direction. \citet{pirogov2009}
carried out further studies of these data through analysis of the cloud's radial density profile.

Our main goal in this work is to study the stellar population of the associated massive 
star cluster, as well as, to investigate the structure of the interstellar 
magnetic field in the direction of the related H{\sc ii} region.
This was done by combining new deep NIR imaging with optical and NIR polarimetric 
data of the related region. The paper is organized as follow: 
Section 2 describes the NIR photometric data and reduction techniques, as well as, the 
polarimetric observational sample. Results and analysis obtained from the 
photometric and polarimetric data are separately exposed respectively in 
Sections 3 and 4. Discussion of these results are introduced in Section 5, and the 
final conclusions are listed in Section 6.

\section{Observational Data}
\label{obsdata}

\subsection{Near-Infrared Photometry}

The raw imaging data for the IRAS 09149-4743 region were retrieved from the ESO
archive, and the observations comprise two separate programs with the following
identifiers: 073.D-0102 (PI Dr. Sergio Ortolani) and 080.D-0470 (PI Dr. Ben Davies). 
The observing missions were conducted during the nights of May 5th 2004 
and February 10th 2008 using the SofI (``Son of Isaac'') NIR camera \citep{moorwood1998}, mounted on the 
NTT 3.58\,m telescope from La Silla/Chile.

In Figure \ref{phot_datasets}a we use a combined 2MASS \citep{skru2006} image of the 
studied region to show the surveyed area. The 2004 survey corresponds to the two small
squares shown in this figure (delimited by cyan dashed lines); one centered on the stellar 
cluster and the other on a nearby control field. This survey will be hereafter denominated 
``Dataset A'', and corresponds to J, H and Ks bands observations using a 
$0.144''$/pixel plate scale, which allows a $2.1\mathrm{'}\times2.1\mathrm{'}$
field-of-view. The observational strategy included 24 expositions at each of the three filters,
corresponding to 12 observations of the cluster field intercalated with 12 observations
of the control field. Moreover, successive observations of the same field were made by jittering
the telescope by a small displacement ($\approx9.5''$, roughly in the NE-SW direction) between 
each exposition. The FWHM values of point-like sources at this observational run are of $\approx1$''.

\placefigure{phot_datasets}

The observational data from the 2008 survey (which we hereafter denominate ``Dataset B'') 
was obtained using the H and Ks bands, with a plate scale of $0.288''$/pixel, allowing a field 
of view of about $4.9\mathrm{'}\times4.9\mathrm{'}$. Four jittering positions 
were used, moving the telescope in the N-E-S-W directions while positioning 
the cluster in each of the four quarters of the frame. Using this method,
a larger field could be mapped covering an area of $7.4\mathrm{'}\times7.4\mathrm{'}$ on the
sky, centered on the stellar cluster. In Figure \ref{phot_datasets}a 
the region associated to Dataset B corresponds to the larger square delimited by a solid line.
Atmospheric conditions during this observational run allowed mean FWHM values 
of $\approx0.7$''.

Images from both datasets were treated by using IRAF\footnote{IRAF is distributed by the 
National Optical Astronomy Observatories, which are operated by the Association of 
Universities for Research in Astronomy, Inc., under cooperative agreement with the National 
Science Foundation.}\citep{tody1986} tasks from the NOAO package to perform the following 
reduction steps: division by a flat-field frame and illumination pattern, correction of bad pixels 
from the frame using a bad pixel mask, correction of the ``crosstalk effect", combination of 
jittered images in order to create sky fields, subtraction of the sky images, combination of 
the sky-subtracted images and alignment of images from different filters. The reduced 
images of Datasets A and B are respectively displayed on Figures \ref{phot_datasets}b 
and \ref{phot_datasets}c. MONTAGE\footnote{Eletronic address: 
http://montage.ipac.caltech.edu/} 
was used to create the $7.4\mathrm{'}\times7.4\mathrm{'}$ field mosaic of Dataset B 
(Figure \ref{phot_datasets}c). 

DAOFIND (from the DAOPHOT package) was used to locate stars peaked with $3\sigma$ 
above the background, and objects missed by this routine were further inserted by 
visual inspection. Astrometric configuration of the objects' world coordinate system was done 
by using a sample of 2MASS isolated objects from the same field. Correction
of the coordinates was achieved with a rms uncertainty of $\approx0.05$ arcsec 
in right ascencion ($\alpha$) and declination ($\delta$), for both Datasets.

PSF photometry was further performed using the algorithms from
the DAOPHOT package \citep{stetson1987}, with the following photometric parameters for 
each Dataset: 
PSF radius of $2.30''$ and fitting radius of $1.30''$ for Dataset A;
PSF radius of $2.59''$ and fitting radius of $0.86''$ for Dataset B.
Several runs of psf fitting and stellar subtraction were performed in order to reveal 
and obtain magnitudes for the very close faint companions from the most crowded areas.

Photometric calibration was achieved by comparing the instrumental magnitudes 
with a sample of isolated, good quality photometry stars (flag ``A") at the same field
from the 2MASS survey. Such comparison is shown in Figure \ref{comparison_completeness}
for both datasets. These diagrams show a quite good correlation, with zero point uncertainties 
of $0.02$ mag in all photometric bands for both datasets.
Individual magnitude uncertainties were computed using a quadratic sum of the 
instrumental errors from the psf photometry and the calibration uncertainty.

\placefigure{comparison_completeness}

Furthermore, approximate completeness limits were derived from the analysis of the histograms, 
also shown in Figure \ref{comparison_completeness}.
It is defined as the point where the histograms deviate from the straight line, which represent
a linear fit to the logarithmic of the number of objects per magnitude bin. The values
obtained for Datasets A and B are
$(\mathrm{J},\mathrm{H},\mathrm{Ks})_{limit}=(18.6,17.4,16.8)$ and 
$(\mathrm{H},\mathrm{Ks})_{limit}=(18.4,17.8)$, respectively.
Considering the above photometric limits of both datasets, an improvement
of 1.6, 2.1 and 2.8 magnitudes have been obtained respectively in J, H and Ks bands, 
as compared to the earlier survey by \citet{ortiz2007}, allowing us to probe deeper into 
the lower-mass stellar population.

In order to check the existence of color dependence terms to be applied to the SOFI photometry,
we constructed diagrams of magnitude differences (between the SofI and 2MASS systems)
as a function of the (J-H) and (H-Ks) colors, as can be seen on Figures
\ref{colordep}a (Dataset A) and \ref{colordep}b (Dataset B, only the H-Ks color).
As can be noticed, there is no need to apply color correction terms to the derived SOFI
photometry.

The photometric results are shown in the Appendix, respectively corresponding to 530 and 2608 
stars in Datasets A (Table \ref{tab_phot_A}) and B (Table \ref{tab_phot_B}). 
The full versions of the tables are available only in the electronic version of this
paper.

\placefigure{colordep} 

\subsection{Optical/Near-Infrared Polarimetry}
\label{observationspol}

The optical and NIR polarimetric data used in this work have been separately obtained with
different telescopes and instrumental configurations. The optical data were collected 
using the 0.9\,m telescope of the CTIO (Cerro Tololo, Chile, operated by the National Optical 
Astronomy Observatory - NOAO) observatory, during February and April/2010; and March/2011. The 
NIR data were obtained using the 1.6\,m telescope of the OPD (Pico do Dias, Brazil, operated 
by Laborat\'orio Nacional de Astrof\'\i sica - LNA/MCT) observatory, during April/2011.

The polarimetric modules used in both telescopes are similar and consist basically of 
a rotatable achromatic half-wave 
retarder followed by a calcite Savart plate and a filter wheel \citep[for a complete description 
of this instrument, see][]{magalhaes1996}. The half-wave retarder can be rotated in steps 
of 22\fdg 5, and one polarization modulation cycle is covered for every 90\degr\ rotation of 
this wave-plate. 
This arrangement provides two images of each object on the detector, which correspond to the 
perpendicular polarizations beams $f_o$ (ordinary) and $f_e$ (extraordinary).
By rotating the half-wave plate by 45\degr\ yields in a rotation of the polarization direction 
of 90\degr . Thus, at the detector area where $f_o$ was first detected, now $f_e$ is imaged 
and vice versa. Combining all four intensities reduces flat-field irregularities. In addition, 
the simultaneous imaging of the two beams allows to observe under non-photometric 
conditions and, at the same time, virtually suppress the sky polarization component.
Besides, if extended interstellar emission from the field contributes with some 
polarization level, this component is also automatically eliminated.
Finally, each set of a polarimetric observation was performed using blocks of
8 steps of the half-wave retarder, which corresponds to a $180^{\circ}$ rotation of the plate.

Reductions were performed in the standard manner using IRAF's routines, for both the
optical and NIR data. Point-like sources were selected from the images using DAOFIND, 
for stars peaked $5\sigma$ above the local background level, with the saturated objects 
removed from the sample. The total counts for each source were computed applying aperture photometry 
using the PHOT routine for several ring sizes around each star.
Objects affected by cosmic rays, bad pixels or superposition of light beams from 
close companions were not used in our analysis (classified as bad polarization values).

   \begin{figure}[t]
   \centering
   \includegraphics[width=0.48\textwidth]{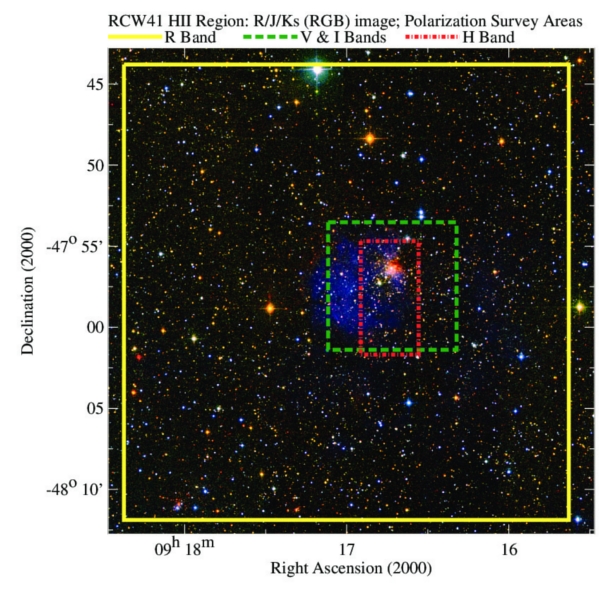}
      \caption{The image above is a RGB combination of the R (DSS), J and Ks (2MASS) bands, 
               showing both the H$\alpha$ extended emission related to the H{\sc ii} Region (blue) and 
               the position of the embedded stellar cluster. The areas mapped by the polarization
               survey are indicated by the boxes related to different spectral bands. The R optical
               band (yellow solid line) has been used to cover a large area ($28'\times28'$) encompassing 
               both the cluster and the H{\sc ii} region, while V and I optical bands (green dashed box) cover a 
               $8'\times8'$ area more focused on the cluster.
               The NIR observations with the H band (red dashed-dotted box) 
               cover a $3.5'\times7.0'$ area towards the cluster and extended
               along the south.
              }
         \label{locpolsurvey}
   \end{figure}

From the difference in the measured flux for each beam, we derived the polarimetric 
parameters using a set of IRAF tasks specifically designed for this purpose
\citep[PCCDPACK package,][]{pereyra2000}. This set includes a FORTRAN 
routine that reads the data files and calculates the normalized linear polarization from a 
least-square fit solution, which yields the degree of linear polarization ($P$), the polarization 
position angle ($\theta$, measured from north to east), and the Stokes parameters 
$Q$ and $U$, as well as the theoretical (i.e., the photon noise) and measured errors. The 
latter are obtained from the residuals of the observations at each wave plate position 
angle ($\psi_i$) with respect to the expected $\cos 4\psi_i$ curve.
In further analysis of the polarimetric data, we adopted the greater value between both estimated errors.
 
Finally, in order to determine the reference direction of 
the polarizer, and to check for any possible intrinsic instrumental polarization, a set
of polarimetric standard stars were observed at each night. The obtained polarizations degree
for the observed unpolarized standard stars proved that the instrumental polarization are negligible 
for both telescopes and instrumentation.

In the case of the optical survey, V, R and I Johnson-Cousins' filters (covering different areas over the studied region) were used. 
In Figure \ref{locpolsurvey}, it is shown a false-color image measuring about $30'\times30'$ that corresponds to a combination 
of the R (DSS=blue), J(2MASS=green) and Ks (2MASS=red) images of the RCW41 region. The extended emission seen in 
blue is probably mostly due to the H$\alpha$ emission of the associated H{\sc ii} region; the cluster objects 
are mainly seen in red and near the centre.
The polarimetric R-band 
survey (yellow box - $28'\times28'$) covers almost the whole area, while the V- and I-bands covers about $8'\times8'$ of the
region centered on the cluster (green dashed line). Each field was observed using 
exposures of 300 sec per half-wave plate position, and in order to cover the entire area of 
the R-band survey, a $3\times3$ frames mosaic mapping was conducted at the CTIO's 0.9\,m 
telescope. The polarized standard stars for the optical survey were selected from the 
catalogs compiled by \citet{tapia1988} and \citet{turnshek1990}, corresponding to 
HD\,298383, HD\,111579, HD\,126593 and HD\,110984.

\placefigure{locpolsurvey} 

The NIR observations gathered at the 1.6\,m telescope of the OPD observatory were performed using 
the H-band of the NIR camera CamIV, which has a $4'\times4'$ field-of-view when mounted
at this telescope. At each position of the half-wave plate, sixty 10 sec images were obtained 
(in order to avoid the detector's non-linearity limit for the brighter sources), jittering the 
pointing by $12$ arcsec between each image in a cross-pattern, summing up to a 600 sec 
of total exposure time. Two fields were mapped in this manner, one centered on the cluster 
and a second one shifted towards de south, therefore covering a $3.5'\times7.0'$ area as 
shown in Figure \ref{locpolsurvey}, by the red dashed-dotted box. The NIR polarimetric 
standards Elias 14, Elias 25 ($\rho$\,Oph) and HD164740, were selected from \citet{whittet1992}, \citet{wilking1980}, and 
\citet{wilking1982}.

The complete polarimetric survey is presented in the Appendix (Table \ref{tab_pol}) with only a portion of the table
given here just to demonstrate its format and content. The complete Table is given only in the electronic version of the article.

   \begin{figure}
   \centering
   \includegraphics[width=0.5\textwidth]{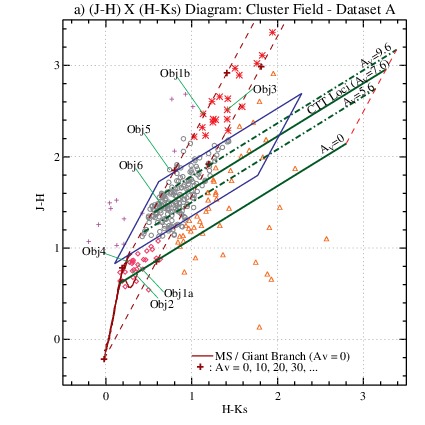}
   \includegraphics[width=0.5\textwidth]{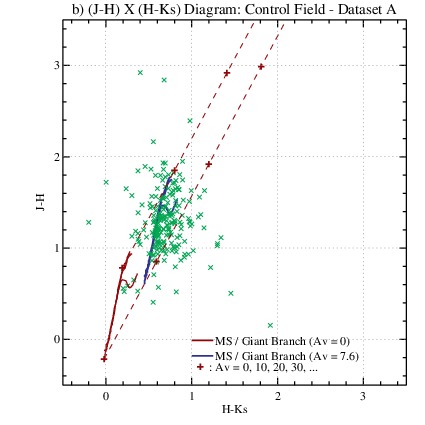}
      \caption{Color-color diagrams for the cluster ({\it a}) and 
               control ({\it b}) areas from Dataset A. Stars marked with Obj1-6 labels 
               are objects previously studied through spectroscopic techniques \citep{roman2009}.
               The red dashed lines represent the reddening band from the standard reddening law, 
               and the dark green lines denote the locus of CTT Stars. Different symbols and colors
               designate different sections from the diagram ({\it a}): the bulk of cluster's stars is represented 
               by gray open circles, with $1.1 < J-H < 2.2$, within the reddening band; also within 
               the reddening band are stars presenting higher (red asterisks, $J-H > 2.2$) and lower 
               (pink diamonds, $J-H < 1.1$) extinction values; purple plus signs and orange triangles 
               respectively designate objects towards the left and right (therefore with color excess) 
               of the reddening band. A small dispersion of $H-K\!s=0.1$ beyond the limits of the reddening 
               band is allowed, to account for photometric uncertainties. 
               All objects from the control area ({\it b}) are marked with green crosses.
              }
         \label{figccdiag}
   \end{figure}

\section{Results and Analysis from the Deep NIR Photometry}

\subsection{Color-Color Diagrams}
\label{secccdiag}

Valuable information regarding the nature of the cluster's stellar population may be inferred 
from $(J-H) \times (H-K\!s)$ diagrams, which are shown in Figure \ref{figccdiag}. They were constructed 
from Dataset A, for objects in the cluster and control areas detected in all three bands.
There are represented the unreddened main sequence (MS), along with the 
giant/supergiant branch locus (shown at the left bottom by the red solid line), as obtained 
from \citet{koornneef1983} and corrected to the 2MASS photometric system using the 
transformation proposed by \citet{carpenter2001}. 
Also, the locus of the Classical T Tauri (CTT) stars \citep{meyer1997} is indicated in the cluster's diagram by the solid dark green line,
with the standard reddening law \citep{rieke1985} being represented (in both cluster and control diagrams) by the red dashed lines.
Different colors and symbols are used to represent objects from different parts of the diagram.

\placefigure{figccdiag}

By comparing the distribution of sources in both diagrams, we can notice that despite the majority of control field stars show 
extinction levels similar to the cluster field stars, the objects from the cluster region present a much 
larger number of highly reddened objects and stars with color excess (respectively red 
asterisks and orange triangles). Furthermore, by analyzing the distribution of the majority 
of points from the cluster's diagram (mainly the gray open circles), we 
note that they are approximately distributed along a strip (indicated by the solid blue 
line polygon) which is roughly parallel to the CTT locus (the dark green line). On the 
other hand, stars from the control field are more ``vertically" distributed in the diagram, 
resembling the reddened Main Sequence locus.
This may indicate that most of the stars from the cluster region are probably 
Pre-Main Sequence (PMS) stars.

\subsection{Mean Visual Extinction}
\label{avcalc}

It is possible to take advantage of the fact that, the bulk of stellar objects from the cluster field 
lay in a distribution which is roughly parallel to the CTT line locus (blue polygon in 
Figure \ref{figccdiag}a), to estimate
the mean visual extinction in the direction of the cluster. We 
initially will assume that the objects located inside the polygon area 
are CTT stars. Therefore, assuming the standard interstellar reddening law
of \citet{rieke1985}, we have applied the de-reddening vector computing the 
visual extinction for each star inside the polygon. The average of these values yields 
a mean visual extinction of $A_{V}=7.6\pm2.0$.
We represent in Figure \ref{figccdiag}a the reddened CTT locus ($A_{V}=7.6$), 
as well as the uncertainty from this computation, i.e., the $A_{V}=9.6/5.6$ CTT loci 
(dark green dot-dashed lines). Although a large scattering of the CTT band is seen, 
we estimate that about $30\%$ of all stars from the cluster field are located 
within these limits. 
Besides, the interstellar extinction is probably spatially non-uniform 
along the cluster's region and contamination from foreground objects is inevitable, 
which contributes to increase the observed scattering.

An independent computation of A$_V$ may be inferred from spectroscopically confirmed O/B stars.
According to the survey by \citet{roman2009}, stars labeled as Obj1a, 
Obj2, and Obj4 in Figure \ref{figccdiag}a, probably belong to the group of cluster's most massive 
objects (respectively B0{\sc v}, O9{\sc v} and B7{\sc v}).
Assuming that they have already reached the MS, we can obtain another value for the cluster's 
mean visual extinction by de-reddening their observed (J-H) and (H-Ks) colors, 
using the intrinsic colors given by \citet{koornneef1983} and the interstellar extinction law
taken from \citet{rieke1985}. Adopting this method, we have found the following A$_V$ values, 
respectively for  Obj1a, Obj2, and Obj4: $7.8\pm1.4$, $7.0\pm1.8$, and $6.4\pm2.3$.
All values are in the range 6-8 magnitudes, 
which is consistent with the computation obtained from the CTT candidates method.

   \begin{figure}
   \centering
   \includegraphics[width=0.5\textwidth]{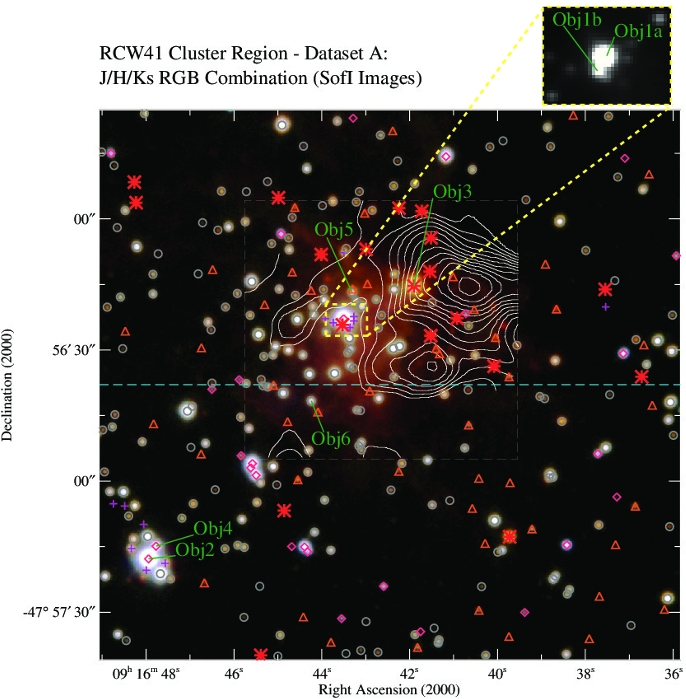}
      \caption{Combined RGB image of the cluster region using J (blue),
               H (green) and Ks (red) images from SofI's Dataset A. Contrast has been enhanced 
               in order to reveal the bright extended emission surrounding 
               the embedded stellar cluster. Stars are marked with colors and symbols 
               related to their location on the color-color diagram (see Figure \ref{figccdiag}a), 
               and objects indicated as Obj1-6 were previously studied through spectroscopic techniques 
               \citep{roman2009}. The white-colored contours are from the HNCO survey by \citet{zin2000}, 
               representing the $10_{0,10}-9_{0,9}$ transition with the intensity levels starting from
               $15\%$ of the peak intensity ($1.5$ K$\cdot$ km s$^{-1}$), raising in steps of $7.5\%$.
              }
         \label{loccc}
   \end{figure}
%

   \begin{figure*}
   \centering
   \includegraphics[width=0.8\textwidth]{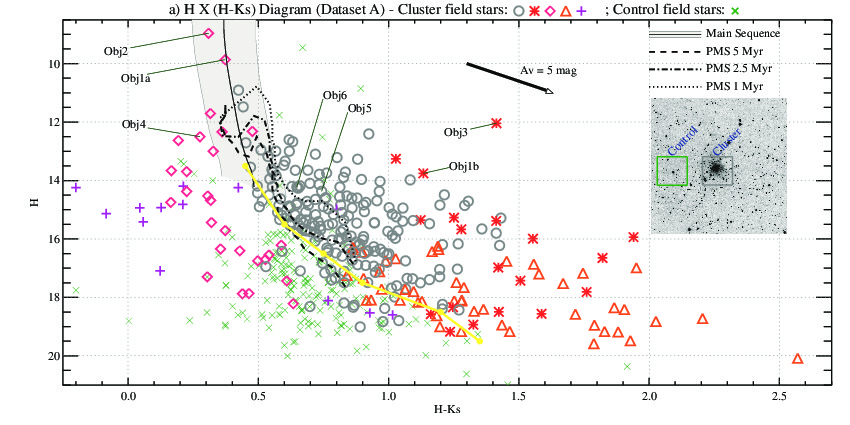}
   \includegraphics[width=0.8\textwidth]{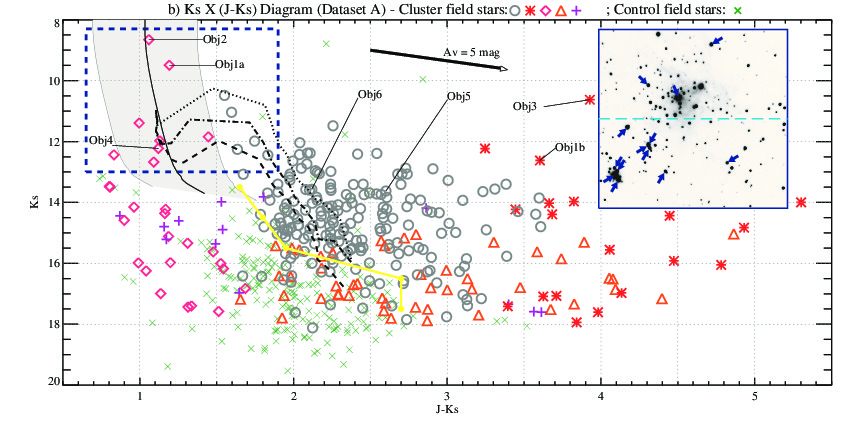}
   \includegraphics[width=0.8\textwidth]{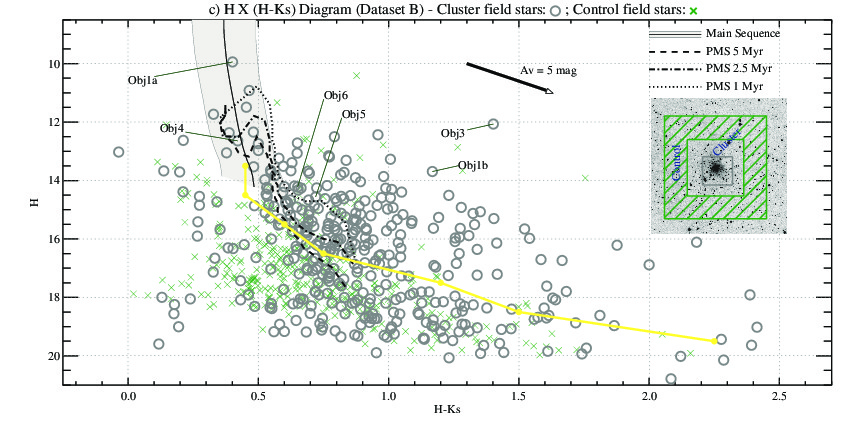}
      \renewcommand\baselinestretch{1.0}
      \caption{
               Color-Magnitude diagrams using Dataset A ({\it a}: $H \times [H-K\!s]$; and 
               {\it b}: $K\!s \times [J-K\!s]$) and Dataset B ({\it c}: $H \times [H-K\!s]$). 
               The small 2MASS images inside diagrams {\it a} and {\it c} show the cluster 
               and control areas used for each Dataset. Stars from the control areas are marked with
               green crosses, while cluster's stars in diagrams {\it a} and {\it b} are 
               marked with colors and symbols according to its location on the color-color diagram 
               (Figure \ref{figccdiag}). All stars from the cluster area in diagram {\it c} are 
               denoted by gray open circles. The MS and several PMS isochrones are indicated, and 
               the yellow line provides a statistical separation between the contaminating field stars 
               and objects from the cluster, as shown in Figure \ref{fighistograms}. 
               In diagram {\it b}, the inset image shows the location of the cluster's objects 
               from the turn-on area (blue arrows), i.e., located within the blue dashed rectangle.
               Obj2 was saturated in Dataset B, and therefore is not shown in diagram {\it c}.
             }
         \label{figcolormag}
   \end{figure*}

\subsection{Spatial Distribution of the Cluster's Stellar Population}
\label{spatialpop}

Considering that the photometric properties of the cluster's area indicate that it is 
composed by stellar populations with different characteristics (embedded PMS stars, 
foreground and background objects, stars with high extinction, etc), it is important to 
study their spatial distribution, relative to the cluster's location. Figure \ref{loccc} shows 
a false color image of the region obtained by combining the J/H/Ks images of Dataset A. 
Each detected star is marked with colors and symbols according to its location on the 
color-color diagram, as defined in Figure \ref{figccdiag}a, with the stars designed as Obj1-6 indicated by the green labels. 
Some very faint stars from the 
field were not marked because they were not detected in all three bands, and 
therefore were not assigned a location in the color-color diagram. 

\placefigure{loccc}

The cluster's region is composed by two distinct structures: the main portion, which 
is associated to an extended emission visible mainly in the Ks band, is formed by a 
number of objects concentrated around the Obj1a/b stars (these two central objects 
present an angular separation of only 1.3 arcsec and are displayed in a close-up 
view at the top of Figure \ref{loccc}); the small sub-cluster region is located 1.1 arcmin 
towards the SE of the main cluster's center, and hosts the Obj2, which is a O9{\sc v} 
star \citep{roman2009}, presumably the most massive object in the field.
According to \citet{ortiz2007}, Obj1a and Obj2 are the best candidates to ionization sources 
of the associated H{\sc ii} region.

The contours shown in the map (colored in white) are from the HNCO survey of massive 
Galactic dense cores by \citet{zin2000}, which covers mainly the central parts of the cluster.
The HNCO molecule is most easily excited radiatively (rather then collisionally) 
inside the molecular clouds' densest regions ($n>10^{6}\mathrm{cm}^{-3}$), being particularly 
sensitive to the far-infrared radiation field from warm dust, and therefore 
is a suitable tracer of regions at the vicinity of massive star-forming sites.
As previously pointed out by \citet{ortiz2007}, the HNCO survey revealed 
an area supposedly swept out by stellar radiation around Obj1a/b, as well as
the presence of a high-density region 30-50 arcsec NW of the central stars
(Obj1a/b). 

We may correlate this information with the location of stars with high 
extinction (marked with red asterisks). These highly obscured objects present $A_{V}$
values in the range 15 - 35, and are most commonly found toward the North of the field 
shown in Figure \ref{loccc}. From a total of 19 high extinction objects, 16 (i.e., 84\% of them) 
are located above the cyan-colored dashed line which divides Figure \ref{loccc} in Northern and Southern 
halves, suggesting an extinction rising gradient in the South-North direction. Furthermore, 
9 of these stars (representing 47\% of them) are located within the HNCO contours, indicating the
expected correlation between the high density medium and the presence of highly obscured stars.
Therefore, although background stars may be present, some of these are most probably highly embedded 
young cluster's sources, and deserve special attention.

There is no clear trending for objects marked with gray open circles or orange triangles,
being scattered along the entire area. However, for pink diamonds (i.e., lower extinction
stars or also massive members of the cluster), there is a weak evidence that an anti-correlation exists
relative to the positions of high extinction stars (red asterisks): 63\% of the pink diamonds (17 among 27 stars) are 
located within the Southern area of the image. Although it is not possible to assert this unequivocally, 
this fact corroborates the idea that a rising extinction gradient occurs in the South-North direction. 
This evidence may support the idea that interstellar material surrounding Obj1a 
is only beginning to be swept out (according to the structure of HNCO contours), 
while the area around the subcluster (which hosts the O9{\sc v} star) have already been 
cleaned due to the action of its strong winds.

\subsection{Color-Magnitude Diagrams}
\label{seccolormag}

In Figure \ref{figcolormag} we show three color-magnitude diagrams for both the cluster 
and control areas. While the control field is probably composed by foreground and background 
main sequence and giant stars, the cluster's field also contains members of the cluster itself.
In order to locate the overlap region, stars from the control area (denoted by green crosses) were plotted in
all diagrams. This is useful to identify stars from the cluster area presenting photometric 
characteristics of contaminating objects.

\placefigure{figcolormag}

Figures \ref{figcolormag}a ($H \times [H-K\!s]$ diagram) and 
\ref{figcolormag}b ($K\!s \times [J-K\!s]$ diagram) were both created with colors and 
magnitudes from Dataset A, using the cluster and control areas as did in 
Section \ref{secccdiag} (the inset in Figure \ref{figcolormag}a shows the cluster's and
control areas overploted in a small 2MASS Ks band image). Points from these two 
diagrams are plotted using the same color/symbol scheme as in Figure \ref{figccdiag},
therefore allowing us to identify the positions of the different groups previously defined in 
the color-color diagram. The zero age main-sequence (for solar metallicity) is represented by a 2MASS JHKs Padova 
isochrone \citep{bonatto2004}, which is denoted by the black solid 
line surrounded by a gray band that accounts for the $\pm2.0$ mag uncertainty in 
$A_{V}$. As a complement, we also plotted three pre-main-sequence (PMS) isochrones 
($t=1$, $2.5$, and $5$ Myr, for solar metallicity) taken from young stellar 
evolution models by \citet{siess2000}.
We used the cluster's distance ($d = 1.3\pm0.2$\,kpc) and the interstellar
extinction as in Section \ref{avcalc}.
Furthermore, an arrow corresponding to an amount of 
5 mag in visual extinction, was also included in each diagram.

Figure \ref{figcolormag}c is a $H \times (H-K\!s)$ diagram built with the H and Ks data 
from Dataset B, which provides a larger area around the cluster (see 
Figure \ref{phot_datasets}). As shown by the inset, we have kept the same cluster's 
area used in Dataset A (Figures \ref{figcolormag}a and b) and chosen a different control 
region, encompassing a wide area at the external borders of the region covered by 
Dataset B (defined by the green cross-hatched). This should allow a more complete
probe of the field stellar population. Since J band data are not available from Dataset B, 
a position in the color-color diagram was not assigned, and therefore, we have simply used 
gray open circles to represent cluster's stars and green crosses for objects from 
the control area.

By analyzing the distribution of the cluster's and control's stars in all three diagrams, we 
notice some striking common features. 
Initially, the majority of the point sources from the cluster's region appear shifted 
from the MS locus, being displaced to its right with higher color values, i.e, 
over the region where the PMS isochrones are present. 
This indicates that the majority of the cluster's sources are probably very young stars
that still have not reached the hydrogen burning phase. However, as would be expected,
some objects from the cluster area are located toward the left part of the diagram, occupying the 
same area as the control field stars, and therefore are probably members of the Galactic field 
stellar population. Also, from figures \ref{figcolormag}a and b, we can see that objects positioned toward the left side
of the ZAMS are mainly those represented by pink diamonds and purple plus signs, i.e., 
lower extinction stars and objects with anomalous colors, therefore indicating that such
objects are likely foreground stars. On the other hand, it is evident that some of the pink diamond brighter stars
(for example, Objects 1a, 2 and 4), are in fact cluster's members that have already reached the MS phase.

   \begin{figure*}
   \centering
   \includegraphics[width=\textwidth]{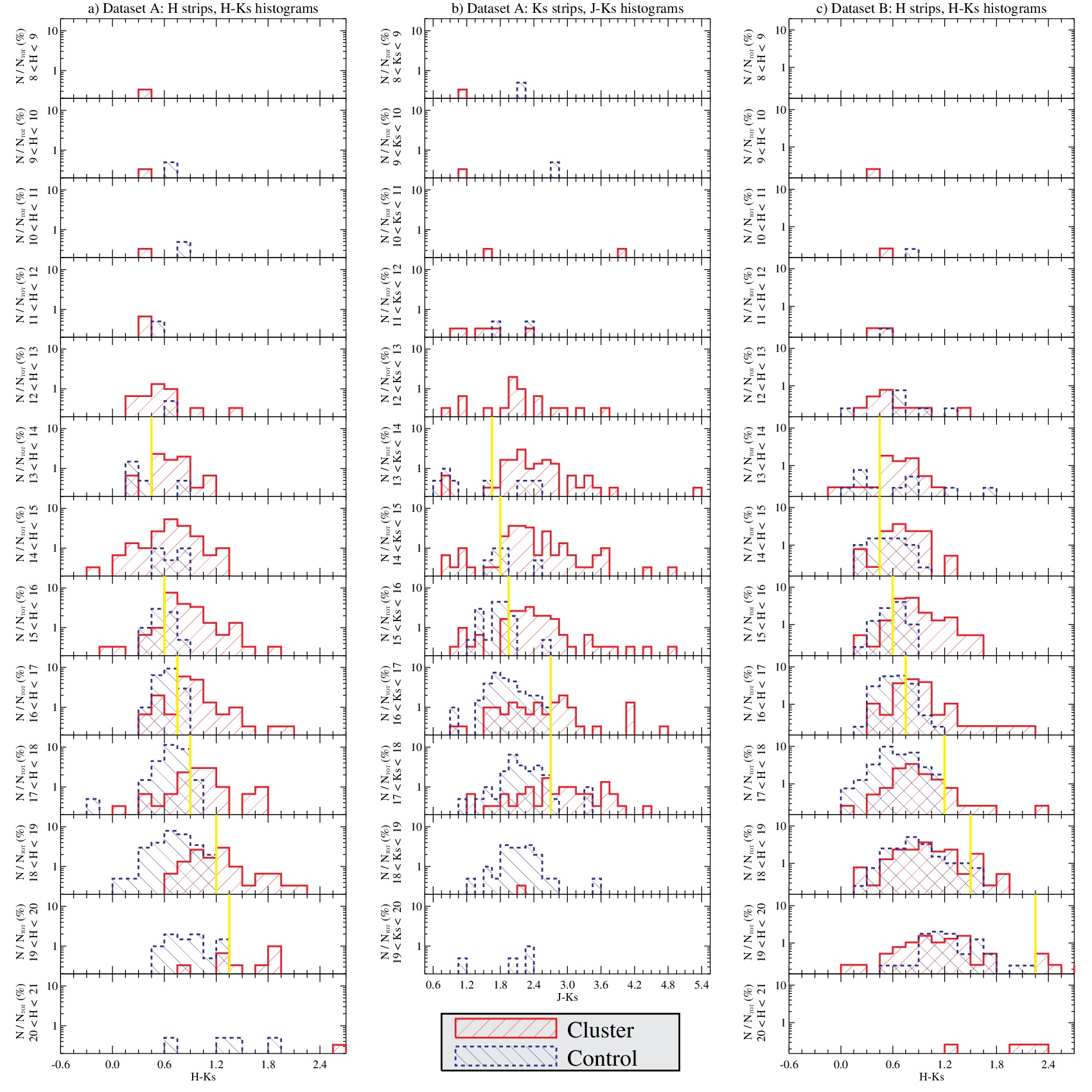}
      \caption{Histograms of color indexes for cluster (red solid lines) and control areas (blue dashed lines), 
               separated by successive strips of 1 magnitude wide. Histograms from parts 
               {\it a}, {\it b} and {\it c} are directly related to the distribution of the 
               stellar population in the color-magnitude diagrams respectively from 
               Figures \ref{figcolormag}a , \ref{figcolormag}b, and \ref{figcolormag}c. Therefore, 
               parts {\it a} and {\it b} are both from Dataset A, respectively representing
               histograms of $H-K\!s$ and $J-K\!s$ in magnitude divisions of $H$ and $K\!s$. 
               Histograms from part {\it c} is from Dataset B, where the $H-K\!s$ color is used 
               in bands of $H$ magnitudes.
               Each value in the histograms were normalized by the total number of stars used 
               in each field.
               Vertical yellow lines are used in several histograms to help identifying contaminating 
               stars from the cluster's region, by separating the overlapping 
               areas between control and cluster populations (at the left of the yellow line) 
               from the population which is mainly composed by cluster's stars (at the right).
              }
         \label{fighistograms}
   \end{figure*}

Although the bulk of the cluster's stars (represented by the circles),
are located amongst the PMS isochrones, a certain number of objects appear 
distributed to the right in the diagrams, therefore presenting large  $J-K\!s$ and 
$H-K\!s$ values. This is probably a combined effect of intrinsic infrared excess from PMS 
stars, non-uniform interstellar extinction 
along the clusters' area, and photometric uncertainties. We specifically notice those stars represented by the orange triangles, 
which are well displaced toward the right showing high color indexes, 
which is an indication of emission from circunstellar disks. Furthermore, these
objects are relatively faint with $H > 16$, therefore probably corresponding to the very low-mass
component of the YSO candidate sample.

Finally, the objects represented by red asterisks (identified as objects within the reddening band, but showing 
high extinction levels) appear presenting some of the highest $J-K\!s$ and $H-K\!s$ 
values. Although some of these stars 
may be background sources, considering the non-uniform and clumped nature of the 
interstellar environment around the cluster (see, for example, the HNCO emission contours 
in Figure \ref{loccc} and the discussion in Section \ref{spatialpop}), it is possible that some 
of them are actually highly embedded cluster's PMS stars. For example, it is highly probable that Obj1b (the 
close companion to the central star Obj1a, and that is marked with a red asterisk) is actually a massive YSO.
More on this subject will be further discussed in Section \ref{objectswithspectra}.

\subsection{The Cluster's Evolutionary Status}
\label{secevolution}

The PMS isochrones from \citet{siess2000} shown in Figure \ref{figcolormag} may be correlated 
with the stellar distribution in the color-magnitude diagrams,
to provide a mean estimate of the cluster's age. However, contaminating field stars from 
the cluster's area should be previously identified, at least in a statistical manner. 
Such identification is based on the overlap areas on the color-magnitude diagrams between 
the cluster and control populations. 

In order to provide a more quantitative recognition
of the contaminating objects, the following method have been adopted: the three color-magnitude 
diagrams from Figure \ref{figcolormag} were divided in bands of 1 magnitude, and to every strip, 
a histogram was constructed, using the color index of each respective diagram. 
The results are shown in Figure \ref{fighistograms}, where histograms 
from parts a, b, and c are respectively related to the color-magnitude diagrams from Figures 
\ref{figcolormag}a, \ref{figcolormag}b, and \ref{figcolormag}c. Each histogram is divided between 
the cluster (red solid lines) and control (blue dashed lines) populations, and bins of 0.15 mag 
were used. The histograms were normalized by the total number of stars used 
in the color-magnitude diagrams of each area, namely: $N_{tot}$(Dataset A, cluster) $= 302$,
$N_{tot}$(Dataset A, control) $= 202$, $N_{tot}$(Dataset B, cluster) $= 384$
and $N_{tot}$(Dataset B, control) $= 396$. Therefore, these diagrams represent the 
percentage fraction of stars within each magnitude strip and color bin. 

\placefigure{fighistograms}

A rough separation between the intrinsic cluster's stars (mainly displaced toward the right of the histograms)
and the field stellar population may be defined by the color index value where the cluster's fractional number 
of stars becomes predominant relative to the control. To every histogram presenting a sufficient number of points, 
this value have been marked with a yellow vertical line. Therefore, it represents a separation criterion between the 
field stars (objects from the left of the yellow line) and the intrinsic cluster's population 
(stars from the right of the yellow line). These values were also indicated in Figure \ref{figcolormag} by the yellow 
connected dots and lines. This provides an appropriate scheme to roughly de-contaminate
the color magnitude diagrams, therefore allowing a more clear analysis of the isochrones' positions
relative to the distribution of possible intrinsic points from the cluster.

Specially when analysing Figures \ref{figcolormag}a and \ref{figcolormag}c (the $H \times [H-K\!s]$ diagrams), we notice that
the yellow separation lines passes exactly between the locus related to $t=2.5$ Myr and 
$t=5.0$ Myr. Moreover, in these two diagrams, the bulk 
of the cluster's points (gray open circles), which are toward the right of the yellow line, seems to 
roughly define a band around the $t=2.5$ Myr isochrone. 

On the other hand, the turn-on point is better defined in the $K\!s \times (J-K\!s)$ diagram 
(Figure \ref{figcolormag}b), as indicated by the blue dashed rectangle, which encompassed both
objects that have already reached the MS, and those that are currently leaving the PMS stage. 
Notice that over the MS region, below the turn-on point for the $t=5.0$ Myr isochrone, 
there is a small gap for cluster's sources with magnitudes in the range from $K\!s=13$ up to $K\!s\approx14$
(from whereon contaminating objects seem to dominate), 
suggesting that there are no stars from the cluster with $t > 5.0$ Myr. 
Furthermore, the Obj4 star, whose distance is consistent with being an intrinsic cluster's source
\citep{roman2009}, is located over the MS, at the midway between the $t=2.5$ and $t=5.0$ Myr isochrones. 

These evidence suggest that the cluster's mean age is between $2.5$ and $5.0$ Myr. Moreover,
these analysis reveal that a sequential star formation scenario may have occurred within this region.
The spatial location of objects from the turn-on area in Figure \ref{figcolormag}b (within the blue dashed rectangle)
is indicated by blue arrows at the inset image in the same diagram. It is evident that most 
of these objects (10 among a total of 13) are positioned closer to (or within) the subcluster, 
instead on the central main cluster area (as would be expected if mass segregation processes had
already taken place). Therefore, a plausible explanation to this evidence may be that the most massive 
objects (located at the subcluster), were formed earlier, subsequently inducing star formation
at the main cluster area. 

This evolutionary view is supported by the fact that the interstellar material 
near the subcluster seems to have already been swept out, while at the cluster area denser clumps 
are still present (see Section \ref{spatialpop}).
If the multi-epoch formation hypothesis is in fact true, the large 
spread of cluster's points on the color-magnitude diagrams towards higher color values 
could also be explained by an intrinsic age spread of the studied stellar population.

\subsection{Analysis of Individual Objects with known Spectroscopic Properties}
\label{objectswithspectra}

The spectroscopic features of several specific objects previously 
studied by \citet{roman2009}, may now be correlated with the objects' position on the 
color-color and color-magnitude diagrams. The same stars from that survey were 
renamed as Obj1-6 and are indicated in Figures \ref{figccdiag}, \ref{loccc}, and \ref{figcolormag}.

Three of the most massive objects from the cluster -- Obj1a, 2, and 4 -- were respectively 
classified as B0{\sc v}, O9{\sc v}, and B7{\sc v}-B8{\sc v} stars. Their location on the 
color-magnitude diagrams (Figure \ref{figcolormag}) are consistent with stars that have 
already reached the MS. 

Obj1b shows spectral lines characteristic of YSO's, and according to \citet{ortiz2007}, 
the set Obj1a+b also indicates features of warm dust, such as infrared emission 
beyond $5\mu$m (probably coming from Obj1b). In fact, this object is located within the 
high-extinction region of the color-color diagram, an indication of its embedded nature.
Furthermore, we notice that its K-band spectrum (see Figure 6 from \citet{roman2009})
presents features which are very similar to the spectrum of T Tau Sb, the third member 
from the T Tau system \citep{duchene2002}: both sources show a large Br$\gamma$ emission
(which is a strong accretion signature), as well as several absorption CO overtones. This evidence 
suggests that Obj1b, the close companion to the massive B0{\sc v} central object (Obj1a), 
is a very young T Tauri star. The presence of this young object near the center of the main 
cluster region provides additional support to the idea that star formation within this area 
have been triggered by the earlier star formation activity at the subcluster, as discussed in Section \ref{secevolution}.

Although the sources labeled as Obj5 and Obj6 were previously classified as late-type field stars,
their position on the color-magnitude diagrams (Figures \ref{figcolormag}a, b, and c) are consistent 
with a classification as T Tauri stars. 
Even though their spectroscopic features may be interpreted as from typical low mass late-type stars, 
observations have been reported of Class I stars with few spectroscopic signs of youth. 
Two examples are the IRAS 03220+3035(S) and IRAS F03258+3105 sources, from the NIR spectroscopic survey 
by \citet{connelley2010}, which were classified as young YSO's, despite presenting K-band spectra only
with some CO overtones in absorption and almost no Br$\gamma$ emission, hence being very similar to 
Obj5 and Obj6. Therefore, if we also take into account their position at the very crowded area 
near the cluster's center (see Figure \ref{loccc}), these sources are most likely T Tauri stars, 
members of the cluster. 

Other interesting properties are related to Obj3, that is a highly reddened source, 
and has NIR colors corresponding to a visual extinction of $A_{V} > 20$, which is consistent
with its spatial location among the HNCO contours (Figure \ref{loccc}). 
Although the previous analysis of its K-band spectrum showed that it could be a typical late-type background star, 
its position in the color-magnitude diagram suggests that it may also be interpreted as
a highly reddened medium mass YSO. A careful analysis of its spectrum show a weak set 
of CO absorption lines, which is usually a feature of low-mass YSO's. However, its 
high NIR luminosity rather suggests a more massive nature. Therefore, a valid explanation 
for the weakness of these lines could be due to veiling of the photosphere and CO bands
by the circunstellar dust emission, as suggested by \citet{casali1995}. In this case, an anti-correlation
of these lines' intensity with the NIR color excess is expected. Although
more observations are necessary to fully characterize this source, these
evidence indicate that Obj3 is likely a medium mass YSO from the cluster.

   \begin{figure}[b]
   \centering
   \includegraphics[width=0.48\textwidth]{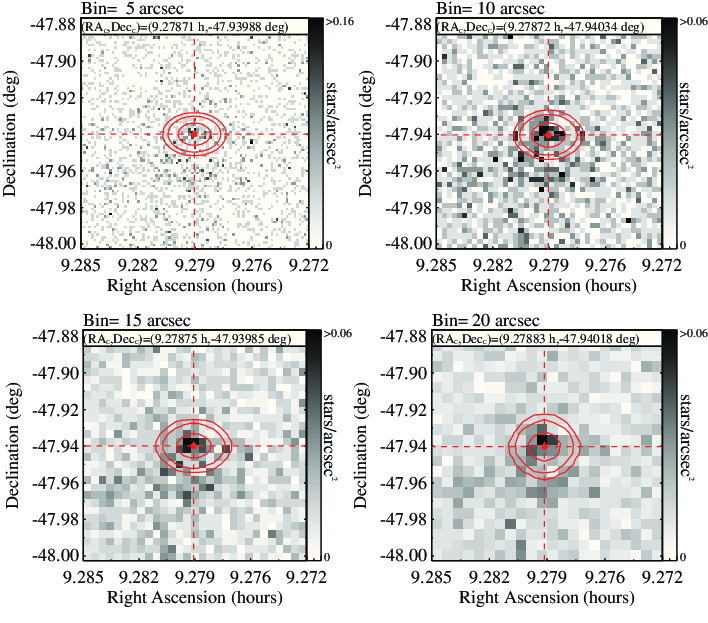}
      \caption{Stellar density distributions to different bin sizes, created using
               the photometric positions of Dataset B. The gray-scale is related to the 
               stellar density, where darker tones represent a higher density, as  
               can be inferred by the scale bars.
               Bi-dimensional Gaussian fits applied to each distribution are represented the red contour lines,
               which are related to the following levels: $0.013$, $0.015$, and $0.030$ 
               stars/arcsec$^{2}$. The center positions computed from the Gaussian fits
               are specified at the top of each diagram, and are indicated as a red bullet.
              }
         \label{cluster_center}
   \end{figure}
%

   \begin{figure}
   \centering
   \includegraphics[width=0.50\textwidth]{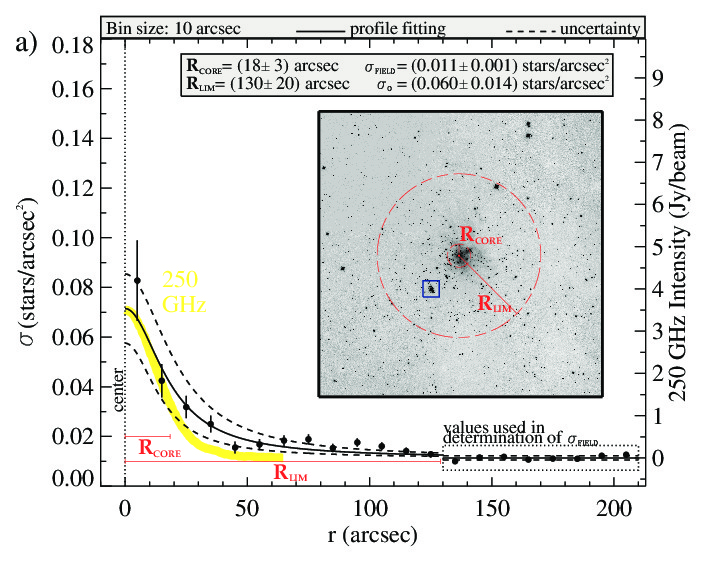}
   \includegraphics[width=0.50\textwidth]{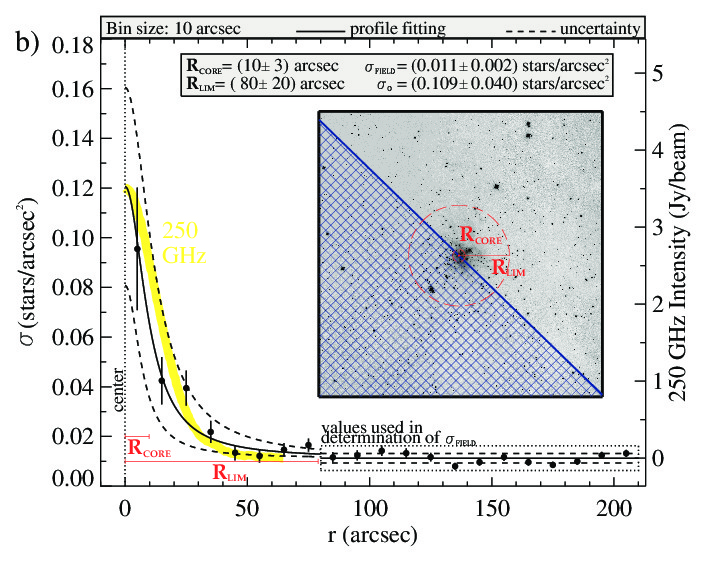}
      \caption{{\it (a)} King profile fitting using concentric rings 
               around the cluster's center position. 
               The values obtained for $\sigma_{0}$, $R_{core}$, $\sigma_{field}$, and 
               $R_{lim}$ are expressed at the top of the diagram, and the spatial size of $R_{core}$ and 
               $R_{lim}$ are indicated as dashed red circles in the Ks image from Dataset B (inset).
               The error bars represent the Poisson uncertainties.
               The $\sigma_{field}$ parameter, which represents the uniform stellar density from the 
               field, is separately computed, by taking the mean value (and the standard deviation) 
               of points with $r > R_{lim}$. 
               {\it (b)} The same King profile fitting, ignoring
               the blue cross-hatched area, which contaminates the RDP due to a stellar density scattering
               toward the South-East. Notice the sharper format of the curve fitting in this case.
               In both diagrams {\it (a)} and {\it (b)}, the thick yellow line represent the 
               radial density profile of the 1.2 mm dust continuum emission (250 GHz) originated from
               the cluster's direction, defining an almost spherical projected core, as described by
               \citet{pirogov2007} and \citet{pirogov2009}. The scale at the right represent the 250 GHz 
               intensity, and the peak of this radial density profile is adjusted to match the peak 
               of the stellar density from the King function, for comparison purposes.
              }
         \label{king_profile}
   \end{figure}

\subsection{Cluster's Center Determination and Radial Density Profile}
\label{radialprofileanalysis}

In order to perform a rigorous analysis of the cluster's Radial Density Profile (RDP),
it is important to determine its mean center. Therefore, we 
have used the Ks image from Dataset B, divided the region into bins of specific widths 
($5$, $10$, $15$, and $20$ arcsec), and computed the stellar density distribution. 
The result is shown in Figure \ref{cluster_center}, where the gray-scale represents 
the number of stars per bin area. 

\placefigure{cluster_center}

The cluster's center position for each bin width was determined through bi-dimensional 
Gaussian fits to the stellar density distributions, represented by the red contour maps 
superposed to each diagram in Figure \ref{cluster_center}. A mean value has been computed, resulting in 
($\alpha_{\mathrm{c}}$,$\delta_{\mathrm{c}}$)=($9^{\mathrm h}16^{\mathrm m}43.5^{\mathrm s}$,$-47^{\circ}56'24.2''$),
with an uncertainty of ($0.2^{\mathrm{s}}$,$0.9''$).
Although the cluster's stars appear slightly scattered toward the South and South-East (near the subcluster), 
a highly peaked concentration appears around $\alpha_{\mathrm{c}}$,$\delta_{\mathrm{c}}$, the cluster's center.

The morphology of the RCW41 cluster's RDP may 
be studied by fittings of a two-parameter King function \citep{king1962,king1966}, which may be approximately described in terms 
of the central density ($\sigma_{0}$) and the core radius ($R_{core}$). This was achieved 
by positioning several circumcentric rings around the computed center, with widths of $10$ 
arcsec each, and computing the stellar density as a function of the distance from the center
($\sigma (r)$). The obtained density profile was fitted by the following equation:

\begin{equation}\label{dens_prof}
\sigma (r) = \sigma_{field} + \frac{\sigma_{0}}{1+(r/R_{core})^{2}}
\end{equation}

The results are expressed in Figure \ref{king_profile}a, where the points indicate the stellar 
densities at each ring, and the  King function fitting is denoted by the solid line, together 
with the uncertainty curves (dashed lines) obtained from the uncertainties of the fitted 
parameters. 
Starting from the center, the stellar density decreases towards larger radius, and the 
point where the cluster's density completely merges with the field density is visually 
identified as $R_{lim}$, i.e., an approximate value for the cluster's limiting radius. 

\placefigure{king_profile}

Note that a bump extending from $r\approx65$ to $110$'' appears in this radial profile, 
which is due to the stellar scattering towards the south-east area (which includes the 
sub-cluster -- small blue box in Figure \ref{king_profile}a). 
In Figure \ref{king_profile}b, we have removed this area from 
the RDP analysis (blue cross-hatched region), in order to provide a cleaner fitting only 
of the cluster's main concentration, i.e., the stellar densities were computed from 
increasing half-rings at the Northwest area of the field. The new King function is much
sharply concentrated toward the center in this case, showing a large decrease of 
$44\%$ in $R_{core}$ and $38\%$ in $R_{lim}$. 
The estimated value of $R_{lim}$ indicates that the control areas used in the 
photometric analysis from the previous Sections, are sufficiently distant from the 
cluster's main structure, therefore avoiding significant contamination. 

\citet{lada2003} pointed out that embedded stellar clusters may be morphologically 
classified as hierarchical-type (with multiple peaks at the density profile, frequently 
found scattered over a large spatial scale) or centrally condensed-type (highly concentrated 
distributions, with relatively smooth RDPs). 
Although the density distributions show that the cluster 
do present some isolated stellar concentrations, the relatively smooth fit of the King function 
indicates that it is probably of the centrally condensed-type. This structural type is a 
common feature of very young stellar clusters, since most of the stars have not yet 
dispersed throughout the field, and are still heavily concentrated around the center.

Such morphology may be correlated with 250GHz observations of the 1.2mm dust continuum 
emission towards this region \citep{pirogov2007}, that defines an almost spherical core in which the cluster
is probably immersed in. Based on these observations, \citet{pirogov2009} computed 
a power law radial density profile of the dust emission, which is shown in Figures \ref{king_profile}a
and \ref{king_profile}b by the thick yellow line (the peak emission was adjusted to match the maximum 
value of the King function). 
One may note that both distributions (from the dust emission and 
from the stellar density) are similar in size, suggesting that the stellar system is still 
embedded within part of the original cloud in which star formation was initially ignited.
Therefore, the presence of such dense interstellar material increases the gravitational
potential and so the physical bounding of the system, aiding to keep the stars 
spatially concentrated. The centrally condensed-type stellar cluster occurs 
because of the interstellar gas and dust from the parental molecular cloud, which 
are still present in the cluster's region, and probably beginning to be swept away due to the 
action of the new-born massive stars.

It is important to point out that in the above mentioned comparison, the centers 
of both the dust emission radial profile and the King function, are not exactly the same, 
being separated by about $\sim 11$ arcsec. Furthermore, \citet{pirogov2009} adopted 
a much larger distance to this object (2.6 kpc), as compared to the value of 1.3 kpc 
spectroscopically determined by \citet{roman2009}. However, the difference is irrelevant
in this case, since we have converted the spatial dimensions to angular sizes in Figure \ref{king_profile}.

   \begin{figure*}
   \centering
   \includegraphics[width=\textwidth]{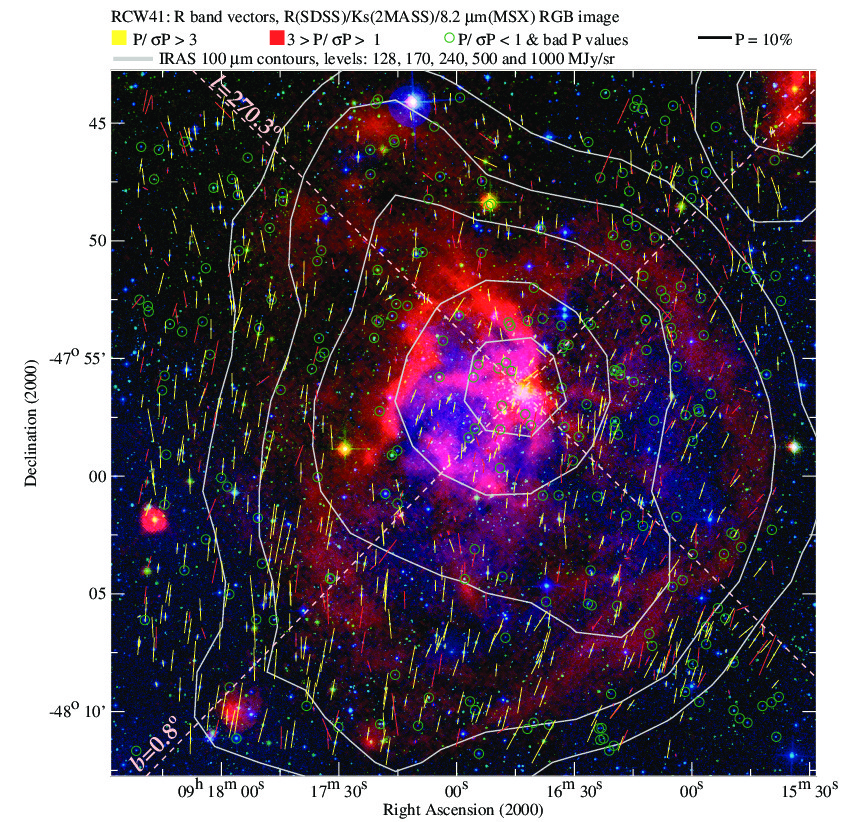}
      \caption{Distribution of R-band polarization vectors at the RCW41 region. 
               The image is a $30\times30$ arcmin RGB combination of the DSS R-band
               (blue), the 2MASS Ks band (green) and MSX 8.2$\mu$m emission (red). 
               Notice the large mid-IR ring or bubble (red) surrounding the blue-colored 
               ionized area. The embedded cluster is located nearly at the center of the region.
               Vectors' sizes are proportional do the polarization degrees
               and colors are related the data quality, as specified above the diagram. Green 
               circles denote objects with poor ($P/\sigma_P < 1$) or bad polarization values. 
               Gray lines are the IRAS 100$\mu$m contours, showing the 128, 170, 240, 500, and 
               1000 MJy/sr emission levels.
              }
         \label{polvecR}
   \end{figure*}
%

   \begin{figure}
   \centering
   \includegraphics[width=0.48\textwidth]{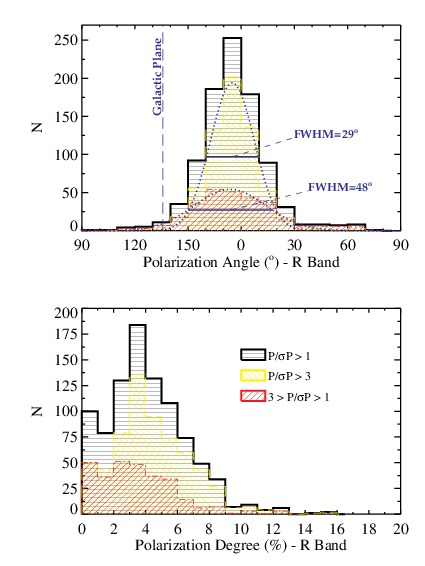}
      \caption{Histograms related to the R-band polarimetric dataset, representing the polarization 
               angle (top) and polarization degree (bottom) distributions. Yellow and red colors
               respectively denote the poor and good-quality data, while the black histograms consider the 
               entire sample presenting $P/\sigma_P > 1$.
              }
         \label{histpolR}
   \end{figure}
%

   \begin{figure}
   \centering
   \includegraphics[width=0.5\textwidth]{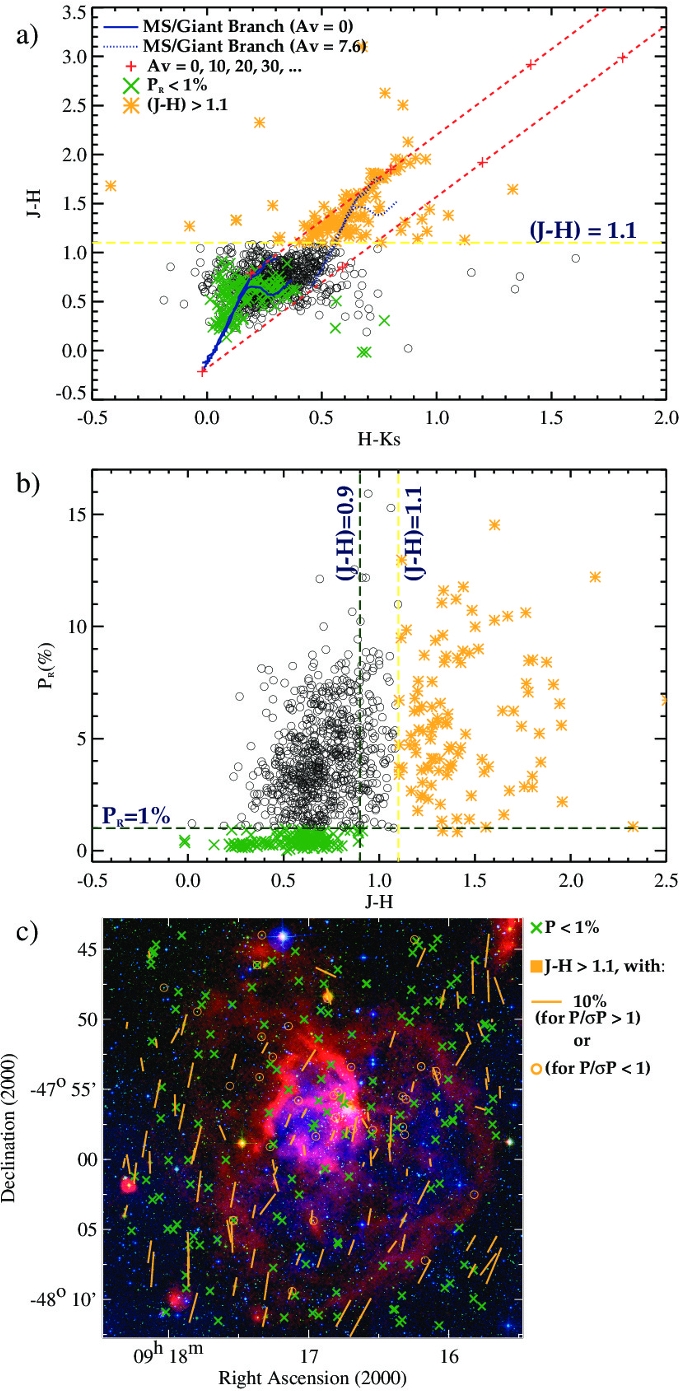}
      \caption{{\it (a)} $(J-H) \times (H-K\!s)$ color-color diagram using 2MASS data for the 
               R-band polarimetric sample; {\it (b)} $P_{R}$(\%) $\times (J-H)$ diagram 
               constructed from a correlation between the R-band polarization data and 
               the 2MASS magnitudes; {\it (c)} Spatial distribution of probable foreground 
               (green crosses) and background or embedded stars (orange vectors, represented as asterisks in diagrams 
               {\it (a)} and {\it (b)}).
              }
         \label{corrpolRphot}
   \end{figure}

\section{Results and Analysis from the Optical/Near-Infrared Polarimetry}

\subsection{The Large-scale Distribution of R-band Polarization Vectors}
\label{polRvectors}

The R-band polarization survey covers a large area around the RCW41 region, 
encompassing both the stellar cluster and the ionized area, as previously shown in 
Figure \ref{locpolsurvey}. A polarization degree and angle were assigned to every 
point-like source detected in the field, and by this manner we have constructed a map 
of the H{\sc ii} region showing the distribution of polarization vectors superposed to each 
star, as exposed in Figure \ref{polvecR}. The vectors' orientations indicate the 
predominant direction of the electric field oscillations related to the stellar light beam, 
and assuming the standard grain alignment mechanisms, the vectors trace the 
sky-projected component of the interstellar magnetic field lines
\citep{davis_greenstein_1951}.

\placefigure{polvecR}

In this image, the size of each vector is proportional to the polarization degree
(a $10$\% sized vector is indicated at the top-right of the graph as reference), 
and the vector's colors are related to the quality of the polarization measurement: 
good quality measurements, with $P/\sigma_P > 3$, are shown in yellow, while 
lower quality data, with $3 > P/\sigma_P > 1$, are shown in red. Green circles denote 
objects with $P/\sigma_P < 1$ or stars with bad polarization values (i.e., affected by 
cosmic rays, bad pixels of the detector or light beam superposition with close companions).
Therefore, all optically detected objects from this survey are marked, either with a vector
or a green circle, according to the above mentioned criteria.

Since the observations were done by integrating the light of all stars from the field 
with the same exposure time, the classification of an individual measurement among 
one of these groups depends on the stellar brightness (which increases the individual 
signal-to-noise ratio) and also on the intrinsic polarization degree from the source 
(therefore, unpolarized sources naturally have larger probability to fall on the 
$P/\sigma_P < 1$ classification).

Considering that many stars are too obscured to be detected 
by the optical survey, only the brightest sources were visible.
Even though, these observations are still useful to study the large-scale properties 
of the magnetic field structure around the H{\sc ii} region, as will be shown in the 
following sections. However, it is important to point out that, 
considering the $1.3$ kpc distance to the cluster, a fraction of these stars 
are foreground objects, and therefore their observed polarization levels
are possibly mainly caused by interstellar components not related to the RCW41 region. 
These properties will be further analyzed in Section \ref{correlationpol2mass}.

The background image used to produce Figure \ref{polvecR} shows interstellar structures 
related to the star-forming region extending into a large area around the embedded 
stellar cluster. The image is a RGB combination of: the DSS R-band image (blue), showing 
optically visible stars and the ionized region (the H$\alpha$ emission is within the spectral 
region defined by the R band); the 2MASS Ks image (green), indicating the location of 
several obscured stars visible only in the near-infrared, including the associated stellar 
cluster, nearly at the image's center; the MSX 8.2$\mu$m mid-infrared emission (red), 
showing mainly the radiation from the warm dust and polycyclic aromatic hydrocarbons (PAHs). 

A striking feature seen in this image is that the mid-infrared emission (red) defines a great 
ring or bubble of warm dust and PAHs that encompasses the ionized area denoted by 
the extended emission in blue. This structure suggests that UV radiation from massive stars 
and high energy photo-ionized electrons from inside the H{\sc ii} region heats the dust grains 
and excites the PAHs from the surrounding environment, generating the ``envelope" type 
of emission. It is not uncommon to find mid-infrared emission of PAHs at the outer edges 
of H{\sc ii} regions, as revealed for example by the {\it Spitzer} GLIMPSE survey of the 
Galactic Plane \citep{churchwell2006,churchwell2007}.

Some mixture between the R-band and mid-infrared emission is seen in the direction of 
the cluster, where several interstellar filaments extend almost radially from it. Such 
morphological structure will be further analyzed in Section \ref{secpolH}.

Another interesting information is provided by the IRAS 100$\mu$m emission which is 
attributed to the colder dust component and is represented in Figure \ref{polvecR} by the gray
contours. Note that the 100$\mu$m emission shows a peak almost centered on the
cluster's area and that the outer contours (especially the $170$ MJy/sr isophote) roughly 
follow the morphology of the hot dust ring from the mid-infrared radiation, suggesting 
the existence of a cold dust ``cocoon" around the entire area. 

Comparing the global distribution of polarization vectors with the direction of the Galactic Plane 
(the $b=0.8^\circ$ line), we note that both directions are not correlated. All-sky 
optical polarization surveys have shown that, particularly along the Galactic Plane,
polarization vectors tend to present an overall horizontal orientation, i.e., parallel to the 
plane \citep{mathewson1970,axon1976,heiles2005}. However, as noted by the latter authors,
the polarization vectors lose this tendency on the magnetic ``poles", located at 
$l\approx(80^{\circ},260^{\circ})$. The VMR is located along one of such ``pole" (the 
RCW41 cluster is at $l=270.3^{\circ}$), and therefore the correlation between polarization 
vectors and the Galactic Plane direction is indeed not expected. 
 
The distribution of polarization vectors presents some characteristics that can be
related with the features of the H{\sc ii} emission region. Note that toward the ionized area, 
and specially near the cluster, the vectors' sizes are small when compared to the outer 
vectors. This evidence indicates a decrease of polarization degree values toward the center
of the region. Furthermore, at the right side of the field, a global bending of the 
vectors' directions is seen both at the area's bottom-right and top-right portions, resembling to 
roughly follow the curvature of the IRAS 100$\mu$m emission contours in these regions. 
Such properties will be further analyzed in Section \ref{polmeanmaps}.

It is important do point out that, since a fraction of the sample is composed 
by YSO's with infrared excess, some of these sources may contribute with a contaminating
polarized emission from the circunstellar disk. However, the large-scale correlated pattern
of polarization angles seen in the map may not be explained by intrinsic polarization. 
The overall characteristics are best described by interstellar polarization due to dust dichroic 
absorption, with some possible scatter in polarization angles due to individual 
disk emission from some sources.

Figure \ref{histpolR} shows histograms of polarization angle and polarization degree divided 
by different $P/\sigma_P$ intervals. The Gaussian fits to the angle histograms show that 
the dispersion (indicated by the FWHM lines) of the $3 > P/\sigma_P > 1$ data (red) is 
larger than the $P/\sigma_P > 3$ case (yellow), an expected trend given the lower 
data quality. However, both distributions are quite well correlated, showing polarization 
angles predominantly in the $160^{\circ}-10^{\circ}$ range. Moreover, by comparing 
the orientations of higher and lower quality vectors in Figure \ref{polvecR}, it may be 
noted that these are quite well correlated locally, indicating that the lower quality data 
may also be included in the analysis.
Polarization degrees varies 
between $0$ and $\sim16\%$, with a large number of stars peaking at $P_{R}\approx2-6\%$.

\placefigure{histpolR}

\subsection{Correlation with 2MASS photometry}
\label{correlationpol2mass}

Since the individual stellar distances for the polarimetric survey are not available, we 
shall take advantage of the 2MASS data to approximately infer the foreground, 
background and intrinsic cluster's populations. Given that the H{\sc ii} region is located 
at $1.3$\,kpc from the sun, it is important to evaluate if the interstellar polarization of 
objects from such distance range are significantly affected by the nearby structures along
the same line-of-sight. Therefore, the goal is to identify, at least statistically, foreground 
objects and analyze their polarization levels in comparison with the typical values from the 
field. 

Figures \ref{corrpolRphot}a and b show respectively the $(J-H)\times(H-K\!s)$ color-color 
and the $P_{R} \times (J-H)$ diagrams using the R-band polarimetric sample 
and the associated 2MASS data. Figure \ref{corrpolRphot}a shows that the bulk of 
polarimetric points present $(J-H)$ values in the range $0.5-1.1$ mag, in contrast with 
the previous photometric analysis (Figure \ref{figccdiag}), where it has been shown that 
stars both from the cluster and control regions are somewhat more reddened. This is 
probably due to a combination of two factors: (1) the earlier analysis was carried out
toward a much smaller area, concentrated on the cluster, where the $A_{V}$ values are
statistically higher; and (2) the polarimetric survey is optically limited, and therefore only 
the brightest sources were detected. 

\placefigure{corrpolRphot}

Figure \ref{corrpolRphot}b shows that, for those objects having $J-H \ga 0.9$, 
there is almost no star presenting $P_{R} < 1\%$.
This effect is emphasized by the vertical ($J-H = 0.9$)
and horizontal ($P_{R} = 1\%$) green dashed lines in this diagram. This is a suggestion that 
objects obscured by the molecular cloud (and therefore located behind or within it) show 
a minimum polarization level equal to $1\%$. As a consequence, stars presenting 
polarization values below this limit are probably foreground objects. Such points 
are marked with green crosses, and define a thick band around the unreddened MS locus 
in the color-color diagram (Figure \ref{corrpolRphot}a), a further evidence that they are 
mainly foreground, not absorbed stars. Moreover, their spatial location on the field is 
uniform (Figure \ref{corrpolRphot}c), as expected for a randomly distributed group of 
foreground objects along the line-of-sight.

We conclude that although a small polarization level due to foreground structures is 
probably included in the stellar light for most of the objects in the observed field, this level 
is small when compared to the typical polarization levels found for the region (which is 
mainly between $2$ and $6\%$, as suggested by the polarization distribution given in
Figure \ref{histpolR}). In fact, \citet{santos2011} have shown that in the directions near 
$l\sim260^{\circ}$ over the galactic plane, polarization degrees in the V band are smaller 
than $1\%$ at least up to a distance of $400$\,pc from the sun. Therefore, it is expected 
that a major fraction of the polarization vectors for the observed field indeed map 
the magnetic field lines within the H{\sc ii} region. 

By analyzing Figure \ref{corrpolRphot}a, we note that a group of points (marked with 
orange asterisks) are located above a small gap (at $J-H \approx 1.1$) within the 
reddening band of the color-color distribution. The $A_{V}=7.6$ MS reddened locus 
(as calculated for the cluster, Section \ref{avcalc}) is superposed to these 
points, showing that these stars are reddened by the molecular cloud, and therefore are 
probably either background or embedded objects. Their spatial distribution is also 
shown in Figure \ref{corrpolRphot}c. Open circles, which comprises the main group of stars 
from the field, have typical polarization degrees above the $1\%$ level for foreground 
objects and typical $J-H$ values below the $1.1$ estimated limit for background and 
embedded stars. Therefore, this group is probably composed by a mixture of foreground, 
background, and mainly embedded objects in the H{\sc ii} region.

   \begin{figure*}
   \centering
   \includegraphics[width=\textwidth]{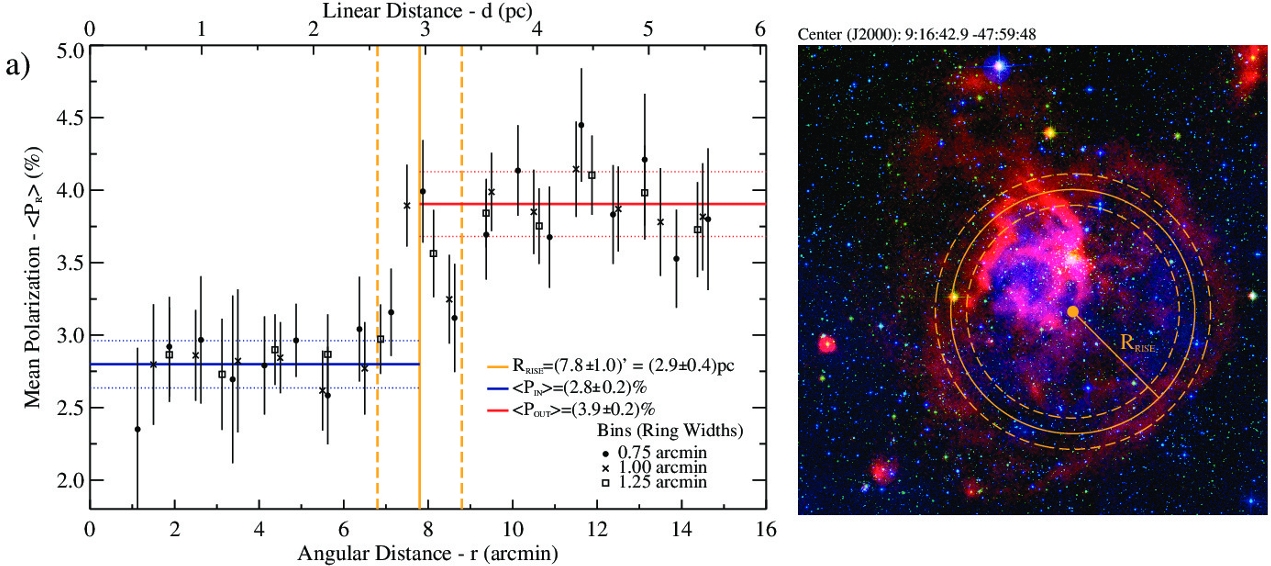}
   \includegraphics[width=0.48\textwidth]{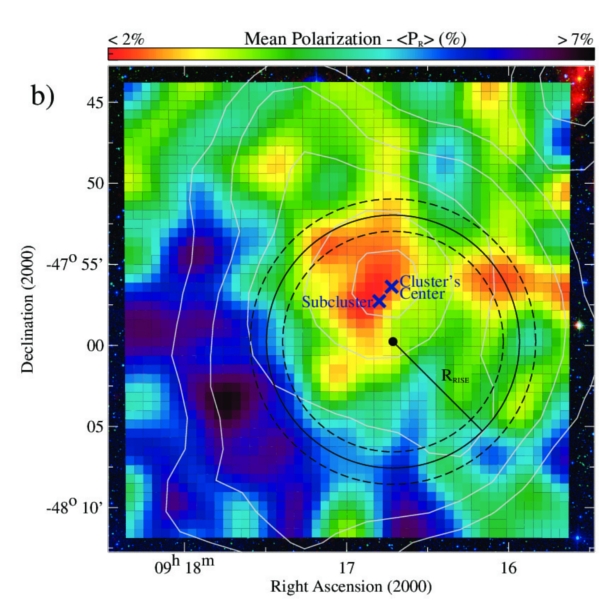}
   \includegraphics[width=0.48\textwidth]{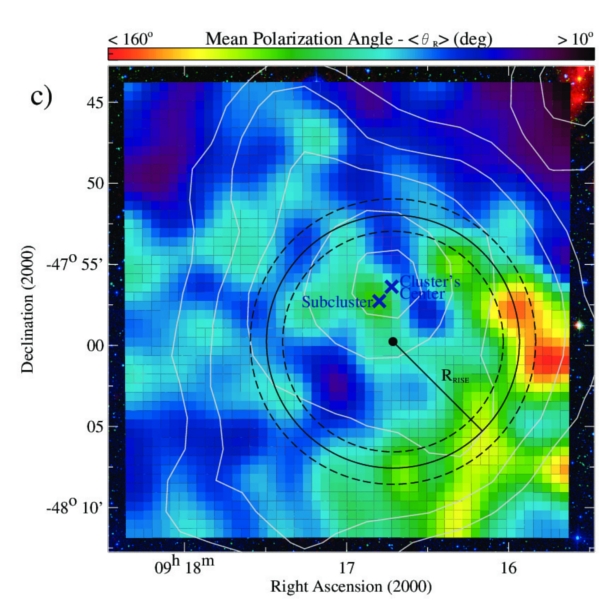}
      \caption{{\it (a)} Analysis of the radial dependence of mean polarization
               degree inside the RCW41 ionized area, where the center has been chosen 
               based on the R-band extended emission (as shown by the blue color 
               at the image toward the right). An abrupt rise in polarization degree 
               is found roughly coinciding with the position of the warm dust/PAHs ring 
               seen at the MSX 8.2$\mu$m mid-infrared emission (red). 
               {\it (b)} Mean polarization map, built by averaging the polarization values 
               of point-like sources into an image of $50 \times 50$ equal divisions. 
               Notice that mean polarization nearby the cluster are smaller than the typical
               values from the field, and a large area at the bottom left following the ridge 
               of the H{\sc ii} region shows an increase of polarization values.
               {\it (c)} Mean polarization {\it angle} image, following the same procedure
               used in part {\it (b)}. A global bending of the sky-projected magnetic field 
               lines is found at the bottom right of the image, also following the rim 
               of the ionized area.
              }
         \label{mapspolthetaR}
   \end{figure*}

\subsection{Large-scale Mean Polarization Degree and Angle Maps}
\label{polmeanmaps}

In order to analyze further the evidence that the degree of polarization is lower inside 
the H{\sc ii} region, we have chosen an approximate center of the ionized area (based on 
the R-band extended emission, represented by the blue color in the image shown in
Figure \ref{polvecR}), and studied the behavior of the mean polarization as a function of this 
radial distance from the center. This analysis is displayed in Figure \ref{mapspolthetaR}a: 
beginning from the selected center position ($(\alpha,\delta)_{J\!2000}$ = 
(9 16 42.9, -47 59 48)), several circumcentric rings of increasing radius were built, and 
inside each ring, the Q and U Stokes parameters from individual objects 
(with $P/\sigma_P>1$) were averaged, therefore allowing to compute the mean 
polarization value associated to that ring and its associated standard deviation. 
Three different ring widths were used in this analysis ($0.75$, $1.00$ and $1.25$ arcmin), and 
the results are identified with distinct symbols in Figure \ref{mapspolthetaR}a. 
Therefore, each mean polarization value was plotted as a function of the angular
distance from the chosen center.

\placefigure{mapspolthetaR}

A sharp increase in mean polarization degree (from $\approx 2.8\%$ to $\approx 3.9\%$) 
was found at a distance of $\sim 7.8$ arcmin from the center of the H{\sc ii} region 
(corresponding to $\sim 2.9$pc, assuming that the cluster is located at $1.3$kpc from the sun). 
Such distance is denoted by an orange circle in the RCW41 image of 
Figure \ref{mapspolthetaR}a. By comparing the position where the polarization degree
increases (with respect to values inside the H{\sc ii} region) with the MSX 8.2$\mu$m 
mid-infrared emission (red), we find a striking evidence that such position roughly corresponds 
to the warm dust and PAHs ``bubble" that encloses the ionized gas.  

Given these evidences, we have attempted to build a ``mean polarization image" of the 
RCW41 region, based on the R-band survey. To achieve this result, we have divided the 
spatial region where the polarimetric data is distributed into $50 \times 50$ 
intervals of equal width. Considering each observation inside a specific division, a mean 
polarization value was computed (only objects having $P/\sigma_P>1$ were used). 
In order to account for cases where no stars were available inside a particular 
division, a smoothing routine was applied, 
returning the same image, but averaged locally by using a small box that runs through the entire
array\footnote{The SMOOTH function from IDL (Interactive Data Language).}. 
The result is shown in Figure \ref{mapspolthetaR}b, where the polarization 
image is superposed on the actual picture of the RCW41 H{\sc ii} region. Redder colors 
correspond to lower polarization levels while bluer colors indicate higher polarization levels, 
as specified by the colorbar given at the top of the diagram.
The location of the cluster (and the associated sub-cluster), as well as of the ring where 
mean polarization is found to sharply increase, are also indicated. 

Note that nearby the cluster (and slightly toward the northeast of it), we find 
an area where polarization mean values are lower, in comparison with typical values 
for the field, presenting $P \la 2\%$ (marked in red). Furthermore, toward the 
left and bottom side of the area (southeast), we find a large region where the polarization 
degree is higher ($P\sim6\%$, marked in dark blue). This region marks the transition border 
between the inner and outer sides of the southeast part of the H{\sc ii} region, and is 
probably responsible for the sharp increase in polarization 
detected in Figure \ref{mapspolthetaR}a. 

Two models may be proposed to explain the decrease of polarization toward the 
ionized region: (1) given that temperature values of the local interstellar medium 
and the radiation field from hot stars are probably higher inside the H{\sc ii} region, 
it is expected that mechanisms of grain alignment are disturbed by the turbulent medium, 
hindering the alignment efficiency and therefore providing lower $P$ values.
Polarization efficiency may also be affected if the intensity of the magnetic field inside 
the photo-ionized area is intrinsically weaker, which may be true if
field lines have been pushed aside by the expansion of the H{\sc ii} region;
(2) dust grains have been swept away from inside the H{\sc ii} region, due to the action 
of intense stellar winds from the newborn stars, therefore resulting in a lower 
dust column density and consequently lower $P$ levels. Obviously, if the second model 
is true, a similar decrease of interstellar extinction as a function of the distance 
from the center would be expected. Further discussion on the distinction between both
models will be carried out in Section \ref{discussion}.

We have also constructed a ``mean polarization angle image", 
using the same procedures described above. The result is shown in Figure 
\ref{mapspolthetaR}c, where yellow and red colors represent vectors slanted 
to the right and blue colors are regions with mean polarization angles slanted
to the left (with respect to the vertical orientation). By this diagram, we are able to 
highlight global changes in polarization angle with respect to the typical values found 
in the field. In fact, we note that toward the bottom right of the image (southwest) , a large 
area marked with green/yellow/red colors indicate a global slanting of the polarization 
vectors toward the right. Furthermore, such regions seems to roughly follow the 
outer contour of the ionized area, indicated by the circle of Figure \ref{mapspolthetaR}a. 
Such inclination of the polarization vectors may also be seen in Figure \ref{polvecR}, as 
pointed out earlier in Section \ref{polRvectors}. 

This structure suggests that the physical processes that generated the H{\sc ii} region, 
followed by the expansion of the ionized area, caused a global deflection of 
the original magnetic field lines in the field. Such deflection of the magnetic field lines around 
the expanding ionization front is most prominent toward the southwest area surrounding 
the H{\sc ii} region, as exposed in Figure \ref{mapspolthetaR}c. 
As an example, \citet{matthews2002} describes another region were a similar distortion of 
magnetic field lines was observed toward NGC 2024, with polarization vectors being 
swept due to the expansion of the associated H{\sc ii} region. 
\citet{tang2009} also described a similar effect toward G5.89, a $\sim0.01$ pc sized 
ultra-compact H{\sc ii} region at a much earlier evolutionary stage ($\sim600$ years), 
already displaying significant distortion of the magnetic field lines as a consequence 
of the expansion of a shell-like structure.

\subsection{Wavelength Dependence of Polarization and Determination of 
the Total-to-Selective Extinction Ratio}

It is possible to take advantage of the polarimetric survey in multiple spectral
bands to analyze specific properties of the dust particles along this region. 
For instance, the grain size distribution causes a direct influence on the 
underlying interstellar law, which is generally assumed to follow a standard behavior.
Therefore such analysis will enable us to directly test this hypothesis, at least 
in the direction of the cluster and nearby surroundings, were V, R, I and H polarization
data overlap (see Figure \ref{locpolsurvey}).

It is well known that an empirical law, known as the Serkowski relation, may be 
used to describe the spectral dependence of the interstellar polarization 
\citep{serkowski1975}:

\begin{equation}
P_{\lambda}=P_{max}\exp{\left[-K\ln^{2}{\left(\frac{\lambda_{max}}{\lambda}\right)}\right]}
\end{equation}

\noindent 
where the typical value of the parameter $K$ is $1.15$,
and $P_{max}$ and $\lambda_{max}$ denote respectively the maximum polarization 
value and the wavelength where such maximum
point is reached. Furthermore, it is also known that when stellar light trespasses different 
dust layers with different grain alignment conditions, it is possible that the resulting 
beam will present a spectral dependence of the polarization angle, usually represented 
as a rotation of $\theta$ with respect to $\lambda^{-1}$ \citep{gehrels1965,coyne1974,
messinger1997}.

\begin{figure}
   \centering
   \includegraphics[width=0.5\textwidth]{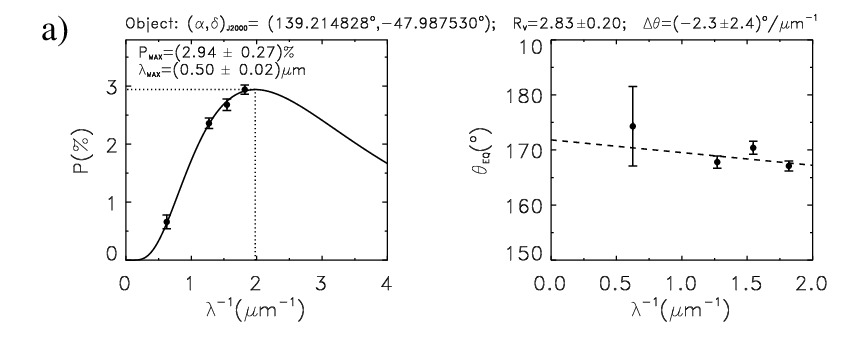}
   \includegraphics[width=0.5\textwidth]{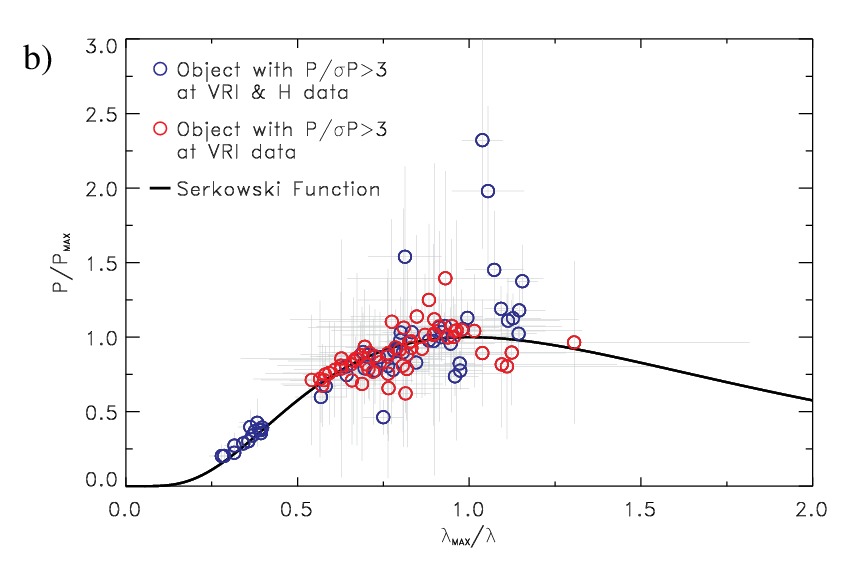}
      \caption{{\it (a)}: Example of the adjusted Serkowski relation ($P \times \lambda^{-1}$) and the 
               wavelength dependence of polarization angle ($\theta \times \lambda^{-1}$) for a particular 
               object presenting V, R, I and H data. The values of $P_{max}$, $\lambda_{max}$, $R_{V}$, and 
               $\Delta\theta$ for this specific case are indicated at the top of the diagrams.
               {\it (b)}: $(P/P_{max}) \times (\lambda_{max}/\lambda)$ diagram, along with the Serkowski 
               curve using points of all available bands for each of the 34 fitted objects. Stars 
               with VRIH or VRI data are distinguished respectively by blue or red open circles.
              }
         \label{serkowskifits}
   \end{figure}

By analyzing the overlap region between the polarimetric surveys in the V, R, I and H
bands, we have selected stars with VRIH or at least VRI observations, where the 
requirement $P/\sigma_P > 3$ was fulfilled in all observed bands. Our polarimetric
sample contains 34 objects meeting these conditions. Values of $P_{max}$, 
$\lambda_{max}$, and a measurement of the rotation of the polarization angle
were obtained for each of these stars. As an example, Figure 
\ref{serkowskifits}a illustrates the results obtained for one of them, i.e., the 
$P \times \lambda^{-1}$ graph (along with the adjusted Serkowski 
curve) and the $\theta \times \lambda^{-1}$ diagram, also showing the fitted values 
for this particular case.

%

   \begin{figure}
   \centering
   \includegraphics[width=0.5\textwidth]{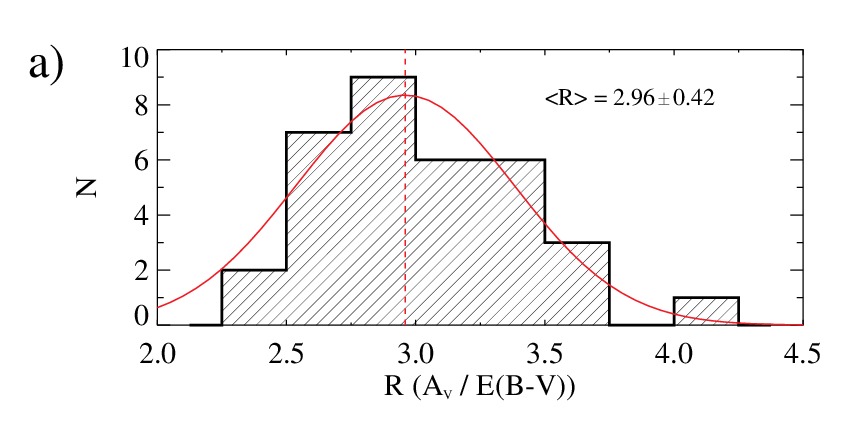}
   \includegraphics[width=0.5\textwidth]{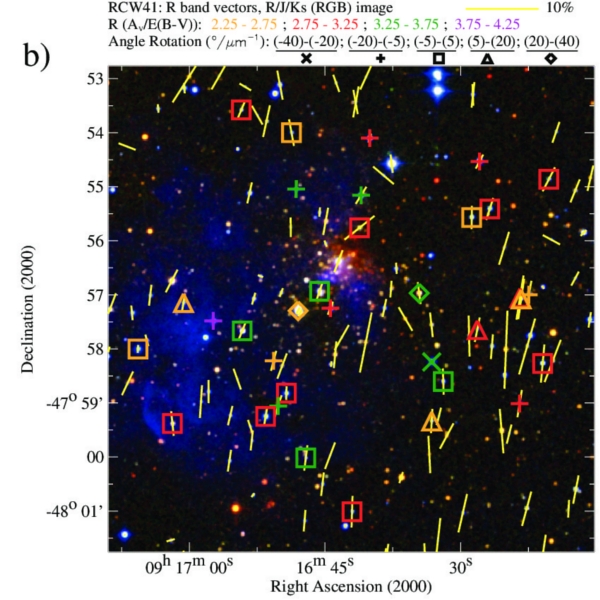}
   \includegraphics[width=0.5\textwidth]{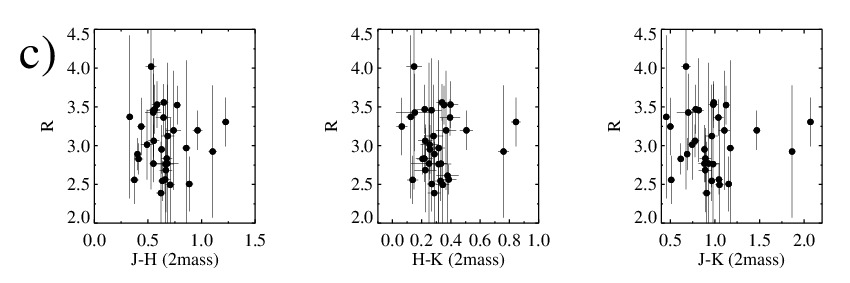}
      \caption{{\it (a)}: Distribution of the fitted values of the total-to-selective 
              extinction ratio (R), along with a Gaussian fit indicating a mean 
              R value of $\overline{R} = 2.96\pm0.42$.
              {\it (b)}: Spatial distribution of the stars studied with the multi-band polarization 
              data (represented by the colored symbols), along with the R-band polarization 
              vectors, both overlaid in a R/J/Ks combined image of the RCW41 region. 
              Symbols denote different angle rotation ($\Delta\theta$) intervals, and different 
              colors represent different strips of the R value, as specified by the caption 
              above the image.
              {\it (c)}: Diagram of the obtained R values as a function of the NIR colors, 
              showing no particular trend or concentration toward lower or higher values.
              }
         \label{Rdetermination}
   \end{figure}
%

   \begin{figure*}
   \centering
   \includegraphics[width=\textwidth]{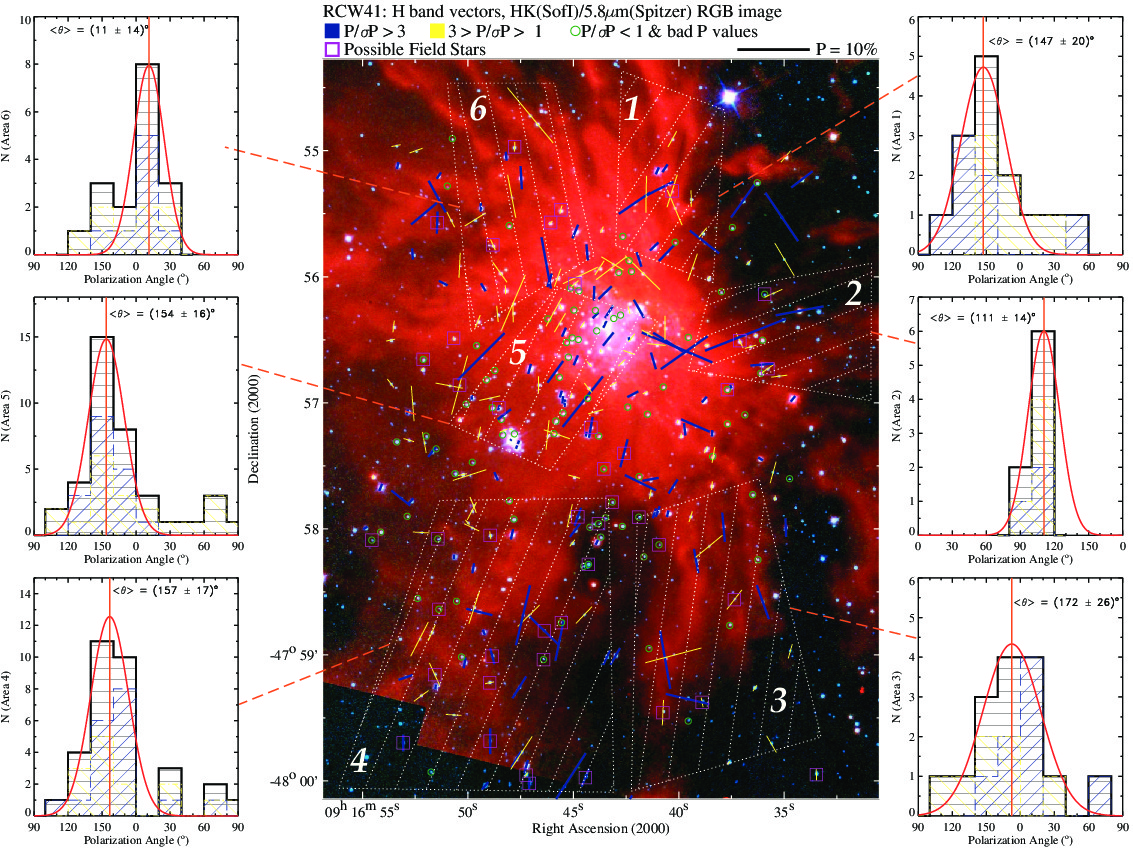}
      \caption{H band polarization vectors map of an area of approximately $3.5'\times7.0'$ 
               encompassing the RCW41 stellar cluster. The image is a RGB combination of the 
               H and Ks images from the SofI's Dataset B NIR survey (respectively blue and green), 
               and the $5.8\mu$m image from {\it Spitzer's} IRAC (red), showing several filaments bended
               toward the cluster. Blue vectors, yellow vectors and green 
               open circles respectively represent data with $P/\sigma_P > 3$ (good quality), 
               $3 > P/\sigma_P > 1$ (lower quality) and $P/\sigma_P < 1$ (also indicating bad polarization
               data). Stars marked with open purple squares are possible contaminating field objects, 
               presenting the following photometric features: $H-K\!s < 0.5$ and $H > 13.5$.
               Histograms denote the distribution of polarization angles for objects within each 
               of the white dotted areas labeled as 1-6 (blue and yellow lines from the histograms 
               also denote respectively good and lower quality data). The peak of each Gaussian fit 
               is defined as the mean polarization orientation from that specific field. This mean
               angle is used to draw a grid inside each selected area, therefore denoting the 
               mean local polarization orientation along that line-of-sight.
              }
         \label{vecpolH}
   \end{figure*}
%

   \begin{figure}
   \centering
   \includegraphics[width=0.5\textwidth]{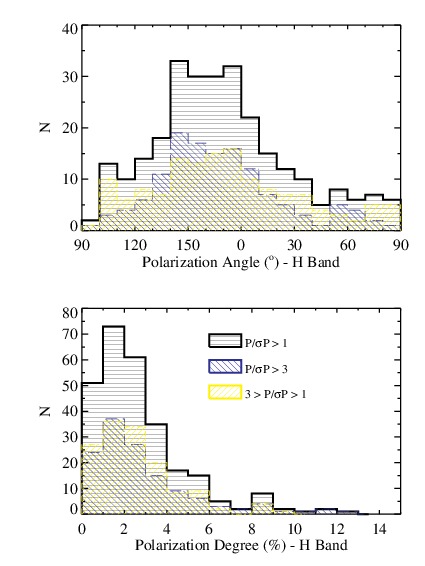}
      \caption{Distributions of polarization angle ({\it top}) and polarization degree 
               ({\it bottom}) for the NIR polarimetric survey in the H band. Blue
               lines are related to the data with $P/\sigma_P > 3$ and yellow lines 
               represent data with $3 > P/\sigma_P > 1$.
              }
         \label{histpolH}
   \end{figure}

\placefigure{serkowskifits}

Figure \ref{serkowskifits}b shows a diagram of $(P/P_{max}) \times (\lambda_{max}/\lambda)$, 
where the individual $P_{max}$ and $\lambda_{max}$ fitted values were used together 
with the polarimetric data for the available bands. Red circles indicate polarimetric data to 
which the fitting was performed only with the VRI bands, while blue circles include also the H 
band. Note that, comparing the points with the position of the Serkowski curve, there is a 
larger dispersion toward lower $\lambda$ values (i.e., toward the V band). This is an 
expected effect since the signal-to-noise ratio of the polarimetric measurement for a particular 
embedded object is supposed to be smaller toward bluer colors. However, the overall 
fitted points forms a band around the Serkowski curve, therefore revealing a quite good 
adjustment.

We estimated the ratio of the total-to-selective extinction, $R_{V}$, by applying the well
known relation between this parameter and $\lambda_{max}$ 
\citep{serkowski1975,whittet1978,whittet2003},

\begin{equation}
R_{V}=(5.6\pm 0.3)\lambda_{max}
\end{equation}

Figure \ref{Rdetermination}a shows a histogram of the obtained $R_V$ values, indicating 
that most of them are concentrated between $\sim2.5$ and $\sim3.5$. By adjusting a 
Gaussian fit to this distribution, the peak value suggests a mean ratio of the 
total-to-selective extinction of $\overline{R} = 2.96\pm0.42$, which agrees with the 
standard value of $3.09$ for the typical interstellar grains. This finding gives support to 
the distance of $1.3\pm0.2$ kpc for the embedded stellar cluster, obtained through 
spectroscopic techniques \citep{roman2009}. Furthermore, it also corroborates the position 
of the PMS isochrones and main sequence locus in the color-magnitude diagrams from the photometric analysis (Sections \ref{seccolormag} and \ref{secevolution}), 
since the standard interstellar law was assumed in those cases. 

\placefigure{Rdetermination}

Figure \ref{Rdetermination}b shows a R(DSS)/J/Ks(2MASS) composed image of the RCW41 
region, with the local R-band polarization vectors distribution, and colored symbols indicating 
the objects from the multi-band analysis. Different colors for the symbols are related to 
different strips of $R_V$ values (divided in intervals of 0.5, as specified above the image), 
and the symbols indicate different intervals of the polarization angle rotation 
($\Delta\theta$, where $\Box$ represent a small rotation level -- between $-5$ and 
$5^{\circ}/\mu$m$^{-1}$ -- while $\times$ and $\diamond$ denote high rotation levels 
respectively of $-40<\Delta\theta(^{\circ}/\mu$m$^{-1})<-20$ and 
$20<\Delta\theta(^{\circ}/\mu$m$^{-1})<40$).

Note that there is no spatial concentration of specific $R_V$ or $\Delta\theta$ values over 
the observed area. Although most of the $\Delta\theta$ values show low rotation levels, a 
great $\Delta\theta$ dispersion is found, with extreme high-rotation values of $-31$ and 
$37^{\circ}/\mu$m$^{-1}$. Along these directions, stellar light has probably passed 
through more than one interstellar layer with different grain alignment conditions. The 
$\Delta\theta$ dispersion is probably a consequence of a turbulent and fragmented 
medium associated to the H{\sc ii} region, with superposing filaments depending on the 
chosen line-of-sight.

Figure \ref{Rdetermination}c shows that there are no particular trend of the fitted $R_V$ 
values to any NIR color index. If stars with color excess due to circunstellar disks also show
an intrinsic polarization level, this should interfere with the fits of the Serkowski relation, 
and therefore with the obtained $R_V$ values. However, these diagrams show no bias of 
the $R_V$ values toward higher color values.

\subsection{The Small-scale Distribution of H-band Polarization Vectors}
\label{secpolH}

The H-band polarimetric survey is focused on the RCW41 cluster (Figure \ref{locpolsurvey}), 
and since most of the objects in this area are highly obscured in the visual wavelengths, 
the NIR survey provides a much larger number of polarization vectors at least 
in a region of a few parsecs around the star-forming locus. 

Figure \ref{vecpolH} shows the distribution of the H band polarization vectors surrounding 
the stellar cluster, superposed on a RGB combination of the H and Ks images (respectively 
blue and green, from SofI's Dataset B), and the 5.8\,$\mu$m image (red) collected using 
the Infrared Array Camera (IRAC)\footnote{The data has been retrieved from 
the NASA/IPAC Infrared Science Archive (IRSA).} onboard {\it Spitzer}. The morphology of 
the mid-infrared extended emission is dominated by several conspicuous filaments 
emanating almost radially from the cluster. These structures resemble the pattern previously
noted in the MSX 8.2\,$\mu$m emission (Figure \ref{polvecR}), but obviously with a much 
higher resolution, allowing the distinction of individual filaments near the cluster.

\placefigure{vecpolH}

Polarization vectors are plotted with different colors according to the data quality
(blue vectors have $P/\sigma_P > 3$ while yellow vectors represent data with 
$3 > P/\sigma_P > 1$), and green circles denote lower quality polarization measurements. 
We have also adopted in this case a criteria to roughly separate contaminating field stars 
from cluster's members, based on the fact that the polarimetric survey in the 
H band reaches an approximate photometric limit at $H=16$. We have found this 
information by correlating the polarimetric data
with the photometric Dataset B and analyzing the H magnitudes distribution for the polarimetric sample.
Therefore, by studying the color-magnitude diagram of Figure \ref{figcolormag}c, we notice that 
up to $H < 16$, the contaminating stars are dominant if $H-K\!s \la 0.5$ and $H \ga 13.5$.
Stars with such photometric characteristics are marked with a purple square in Figure \ref{vecpolH}, 
indicating that these are possibly field stars. A consistent trend in this diagram is that the 
number of field stars within the cluster is small, compared to the outer regions of the image, 
where this number increases.

The first striking evidence found in the distribution of polarimetric vectors is that the polarization 
angle dispersion is much larger than in the R band large-scale survey. This is probably an indication 
of the existence of a largely turbulent medium near the cluster. In fact, if we analyse the histogram 
of polarization angles for the entire H band sample (Figure \ref{histpolH}, top panel), we can see 
that the entire range of possible angles is covered. However, a peak is reached at similar angles 
as in the large-scale distribution, i.e., at $\theta\sim 160^{\circ}$ suggesting that even in smaller 
scales, there is an underlying trend to follow the global scale pattern. Furthermore, no major 
differences are noted between the distributions of the good quality ($P/\sigma_P > 3$, blue) and 
the low quality data ($3 > P/\sigma_P > 1$, yellow). The polarization degree distribution 
(Figure \ref{histpolH}, bottom panel) shows that the majority of stellar objects present $P_{H}$ 
values within $1-4\%$.

\placefigure{histpolH}

Even when considering the large angle dispersion, we find that along even smaller regions 
within the area of Figure \ref{vecpolH}, some local trends may be unveiled, specially toward 
the interstellar filaments seen in $5.8\mu$m extended emission.
In order to reveal these underlying features, we have selected six sub-areas,
mainly encompassing the most prominent filaments, as indicated by the white dotted 
quadrangles marked with the 1-6 labels. Considering the polarization vectors within each of 
these areas, the associated histograms of polarization angles were built, being that objects marked 
as possible contaminating field stars were removed from the analysis. The distributions are 
also shown in Figure \ref{vecpolH}, linked to each respective area, and the division between 
good quality ($P/\sigma_P > 3$) and low quality data ($3 > P/\sigma_P > 1$) are indicated 
respectively by the blue and yellow histograms.
Gaussian fits were applied to each distribution, and the peak of the adjusted function corresponds 
to the main trend of the polarization angles within that area. By using this main value, 
a white dotted grid was drawn within each area, therefore corresponding to the local predominant 
polarimetric orientation.

We note that along the majority of the chosen areas, the main polarization direction
seems to be consistent with the direction of the associated filament. Such trend is specially evident
toward areas 2, 3 and 4. Specifically toward area 2, although only eight vectors contribute 
to the histogram, all of them are oriented parallel to the filament, presenting polarization
angles around $\approx 110^{\circ}$. Such orientation is almost perpendicular to the main trend 
for the large-scale pattern, indicating a clear correlation with the local filament structure. 
Toward areas 1 and 6, a small difference seems to exist between the main polarization angle 
and the filament direction, although the orientation pointing in the direction of the cluster 
is evident. The existence of substructure or superposed filaments with different orientations along 
the line-of-sight may contribute to this displacement. 
Over area 5, which encompasses both the subcluster and a part of the main cluster region, 
an orientation pointing toward the cluster is also evident.

We conclude that despite the large angle dispersion, magnetic field lines on a small scale 
seem to be deflected inwards, pointing toward the cluster and probably connected to the 
direction of the large $5.8\mu$m filaments from the {\it Spitzer} survey. The polarization angle dispersion
is indeed expected to exist in such a dynamical star forming region, where strong stellar 
winds from massive objects and possibly outflows from newborn lower mass stars, should 
play an important role as sources of turbulence to the surrounding interstellar medium.
As a consequence, grain alignment efficiency and magnetic field lines are supposed to be affected, 
therefore characterizing a higher dispersion of the polarization orientations.

\section{Discussion}
\label{discussion}

\subsection{Comparison of the magnetic field structure with theoretical studies 
on H{\sc ii} regions' expansion}

Simulations conducted by \citet{peters2010} describes the collapse of a massive 
molecular cloud followed by the formation of a small stellar cluster, which ionizes the accretion 
flow and causes the H{\sc ii} region to expand. \citet{spitzer1978} first described the evolution 
of a photo-ionized region around an ionization source in a uniform medium, and provided an 
analytical solution to the expansion of the H{\sc ii} region. However, several physical mechanisms, 
such as heating and cooling, turbulence, the effects of ionizing and non-ionizing ultraviolet 
radiation, X-rays, and magnetic fields, must be considered in order to provide a realistic view of 
the expansion process.

\citet{arthur2011} have carried out simulations of this type considering the effects of 
a previously magnetized medium. Comparison of these simulations on non-magnetized 
and magnetized environments provide similar results in terms of the general morphology 
of the H{\sc ii} regions, therefore suggesting that the presence of magnetic field lines does 
not cause a significant impact in providing resistance to the expansion of the ionized area, 
although this influence is most evident when considering the morphology of 
small-scale features, such as globules and interstellar filaments. 

On the other hand, such expansion is expected to cause a large impact on the overall 
shape of the magnetic field lines. The main result from these simulations have shown that
after the expansion of the H{\sc ii} region, a dual behavior of the field's morphology
may be noticed when comparing the inside and outside regions of the ionized area: 
outside the H{\sc ii} region (i.e., in the location of the neutral gas, following its borders), 
magnetic field lines tend to lie along the ionization front, forming a ring around the 
ionized area; inside the H{\sc ii} region (i.e., within the ionized area), magnetic field lines
are expected to lie perpendicular to the ionization front, therefore pointing towards 
the ionization source.

These qualitative predictions are consistent with the results of the large and small-scale
polarimetric surveys of the RCW41 region introduced in Sections \ref{polRvectors}, 
\ref{polmeanmaps}, and \ref{secpolH}. The R-band survey have revealed that magnetic 
field lines are bended along the borders of the H{\sc ii} region toward some areas 
(Figures \ref{polvecR} and \ref{mapspolthetaR}), which is consistent with the interpretation of 
magnetic field lines being pushed aside due to the expansion of the region, therefore lying 
along the ionization front on the neutral areas.

Moreover, our H-band polarimetric survey showed that, although a 
large angle dispersion is evident over the entire area, the mean polarization direction 
resembles a radially oriented field toward the cluster, roughly following the direction 
of several filaments from the mid-infrared {\it Spitzer} image. This is consistent with 
the theoretical prediction that inside the ionized area, field lines are expected to 
lie perpendicularly to the ionizing front. Assuming that the main ionizing sources in
the region are Objects 1a and 2 (from Figure \ref{loccc}), as suggested by 
\citet{ortiz2007} and \citet{roman2009}, magnetic field lines should be pointing 
toward these targets, which is indeed obtained from the polarization vectors mean 
orientation (which points mainly toward Obj1a). 

Such radial pattern is not noticed from the large-scale R-band map, although 
a large portion of this survey is composed by the ionized region, inside the H{\sc ii} region. 
Furthermore, a perfect ring of polarimetric vectors is obviously not formed along the borders 
of the ionized area, as predicted by the simulations of the magnetic field lines. 
However, it is important to point out that the polarimetric survey is not a simple 
sky-projected slice of the H{\sc ii} region. Rather, it represents the  
integrated effect of linear polarization due to different interstellar 
structures superposed along the line-of-sight.
For example, if we imagine the H{\sc ii} region roughly as an expanding sphere, 
such expansion will be noticed spatially in the sky, but will also occur in the line-of-sight
direction. The expansion of the outer borders will push and pull interstellar 
material in our direction, causing the observations of the ionized region to be composed by the 
sky-projected, integrated polarizing effect due to the outer borders and the ionized region
(if the observed star is located behind the H{\sc ii} region). Therefore, the large-scale observations 
of the ionized region should account for this effect, and the radial pattern is not expected to be 
detected.

However, even when considering such effects, an indication of the dual behaviour of the polarimetric
vectors' orientation inside and outside the ionized area is certainly noticed when analysing
the R-band and H-band polarization data. These observations give support to the 
results from the simulations, which provide similar qualitative predictions, at least
when studying the specific environment of the RCW41 region.

\subsection{The origin of the low polarization values within the H{\sc ii} region}

   \begin{figure}
   \centering
   \includegraphics[width=0.5\textwidth]{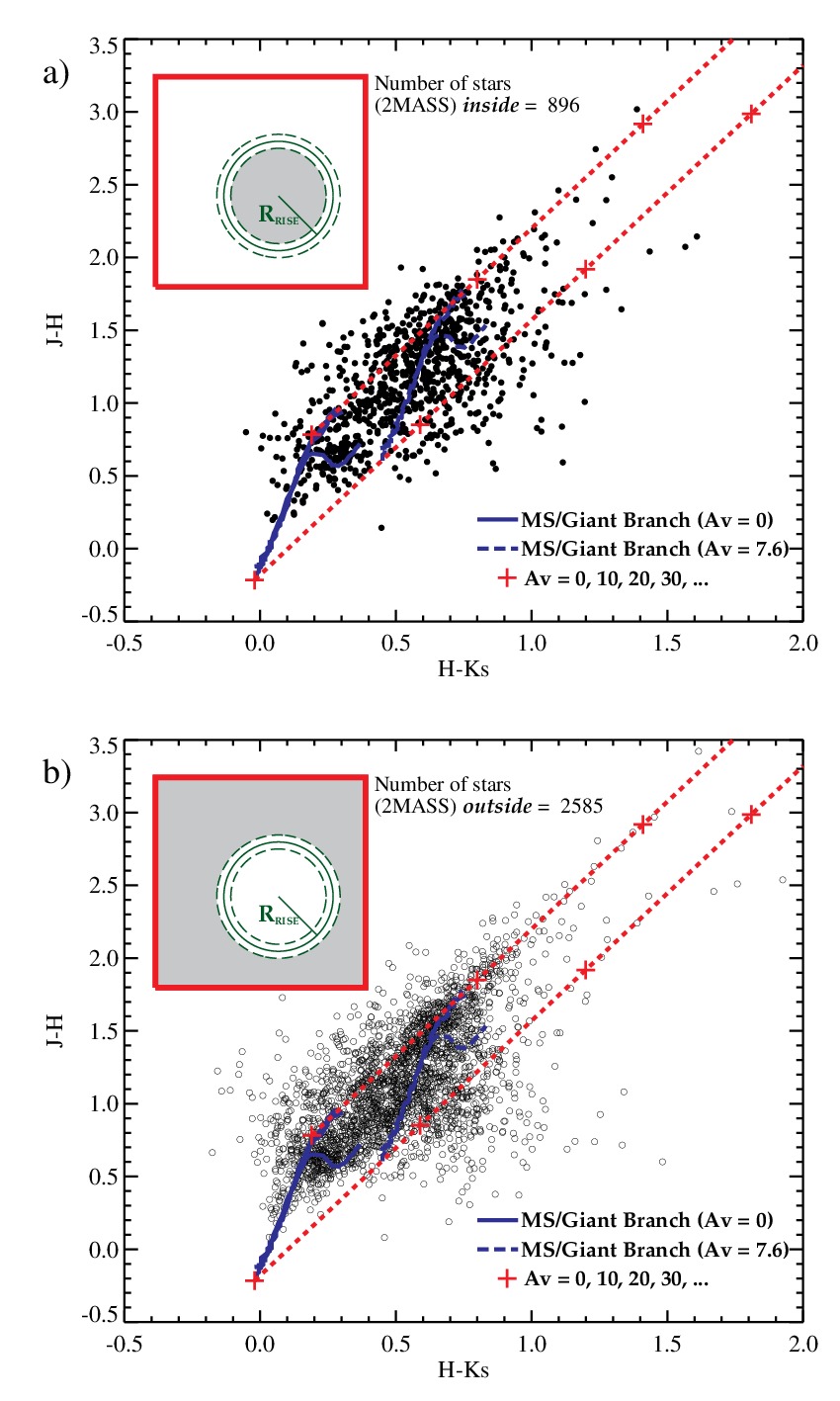}
      \caption{$(J-H)\times(H-K\!s)$ (color-color) diagrams using 
               data from 2MASS, for two separate regions along RCW41
               (as indicated by the red boxes): inside ({\it a}) and outside ({\it b}) 
               the H{\sc ii} region. This area is defined as being delimited by the 
               radius of $R_{rise} = 7.8\pm1.0$ arcmin from the center, where
               a sharp rise in R-band polarization degree is detected, as previously shown 
               in Figure \ref{mapspolthetaR}.
              }
         \label{inout}
   \end{figure}

In Section \ref{polmeanmaps} we have seen that an overall lower 
R-band polarization degrees are detected inside the H{\sc ii} region, specially near 
the cluster. 

In order to distinguish between the two possible mentioned causes responsible for such
effect, two color-color diagrams have been built using 2MASS data, showing the distribution 
of stars inside (Figure \ref{inout}a) and outside (Figure \ref{inout}b) the limits defined 
by the sharp rise in polarization degree ($R_{rise} = 7.8\pm1.0$ arcmin), which roughly coincides 
with the limits of the H{\sc ii} region. A larger number of stars is found outside the region, 
which is obviously due to the wider chosen area. However, when analyzing the distribution of 
stars along the reddening band (which gives an estimate of the interstellar extinction to each object), 
we notice that there are no major differences between both diagrams. In both areas, 
the majority of stars show a roughly uniform distribution of extinction values between
$A_{V}=0$ and $A_{V}\approx 12$. Besides, we know that some concentrations mainly near 
the cluster (which comprises only a small area within the H{\sc ii} region) may 
present larger mean local $A_{V}$ values (as discussed in Section \ref{spatialpop}).
This analysis shows that no global decrease in interstellar extinction is 
detected inside the area.

\placefigure{inout}

Therefore, it seems that the observed decrease in polarization degree inside the H{\sc ii}
region is probably caused by a lower polarization efficiency of the local interstellar
medium. This may be explained by an intrinsic suppression 
of the alignment of dust grains with respect to the interweaving magnetic field lines, 
which is expected given the dynamical and turbulent conditions of such interstellar environment.

Furthermore, in the simulations performed by \citet{arthur2011}, it was shown that
the intensity of magnetic fields in the photo-ionized area is lower ($B=0.1-10\mu$G), 
when compared with the neutral areas from the border of the H{\sc ii} region ($B=10-30\mu$G).
This is probably explained due to the fact that the expansion of the ionized region
causes the magnetic field lines to be pushed aside and therefore removed from inside the 
area, weakening the magnetic field in this location. If this is the scenario occurring in  RCW41, a 
further depletion of polarization efficiency would be expected, contributing to the lower $P$ values 
detected near the cluster.

\subsection{The turbulent component of the magnetic field toward RCW41}

   \begin{figure}
   \centering
   \includegraphics[width=0.5\textwidth]{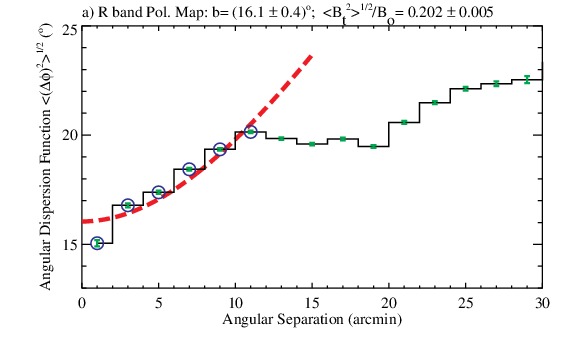}
   \includegraphics[width=0.5\textwidth]{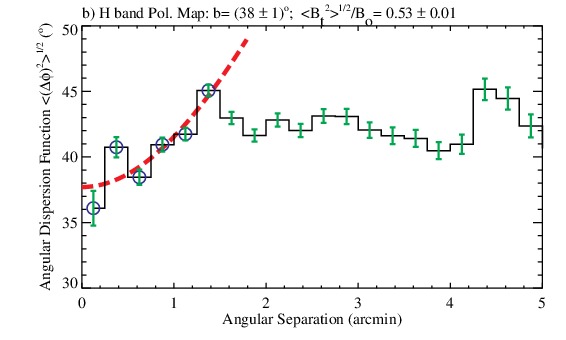}
      \caption{Angular dispersion functions (ADF) constructed by using the 
               polarization angles from the R band {\it (a)} and the H band {\it (b)}
               polarimetric datasets. Each bin (respectively of widths 2 and 0.25 arcmin) 
               have already been corrected due to the mean measurement uncertainties.
               The red dashed line denote the curve obtained 
               through the linear fit of the SF, by using the first six points on each 
               case (blue circles). The turbulent contribution $b$ is determined by the interception of this 
               curve at $l=0$, being subsequently used to compute the ratio of the turbulent 
               to the large-scale magnetic field strength, which are shown at the top
               of each diagram.
              }
         \label{turbulence}
   \end{figure}

The polarimetric dataset toward RCW41 is entirely suitable to apply a method 
which aims on estimating the turbulent contribution to the magnetic field lines, 
relative to its uniform component \citep{hildebrand2009}. It is based on the 
assumption that the interstellar magnetic field $\mathbf{B}(\mathbf{x})$ is 
composed of a large-scale structured field $\mathbf{B_{0}}(\mathbf{x})$, together 
with a turbulent component $\mathbf{B_{t}}(\mathbf{x})$. Initially, the 
procedure consists basically in computing the second-order structure function
of the polarization angles ($\langle\Delta\Phi^{2}(l)\rangle$, hereafter SF), which is defined as the average value of the 
squared difference in polarization angles between two points separated by a distance
$l$ \citep[see Equation (5) given by][]{falceta2008}. The square root of this quantity 
is denominated angular dispersion function (ADF), and provides information on the 
dispersion of polarization angles as a function of the length scale toward a specific 
interstellar area.

\citet{hildebrand2009} pointed out that within the range $\delta < l \ll d$ (where 
$\delta$ and $d$ are the correlation lengths respectively characterizing $\mathbf{B_{t}}(\mathbf{x})$
and $\mathbf{B_{0}}(\mathbf{x})$), the SF is composed by: a constant turbulent contribution (denoted by $b$);
a smoothly increasing contribution with respect to the length $l$, as expected from the
uniform field (the slope of this linear behaviour is represented by $m$); and 
an almost constant contribution due to measurement uncertainties ($\sigma(l)$). 
Being statistically independent, these factors must be added quadratically, leading to

\begin{equation}
\langle\Delta\Phi^{2}(l)\rangle \simeq b^{2} + m^{2}l^{2} + \sigma^{2}(l)
\label{strucfuc}
\end{equation}

The ratio of the turbulent to large-scale magnetic field strength is related to the 
quantity $b$ through the following expression:

\begin{equation}
\frac{\langle B_{t}^{2}\rangle^{1/2}}{B_{0}} = \frac{b}{\sqrt{2-b^{2}}}
\label{btb0}
\end{equation}

Therefore, by constructing the SF (and computing the measurement uncertainties $\sigma(l)$) 
using the polarization angles, it is straightforward to apply a linear fit with Equation \ref{strucfuc}
and determine the $b$ factor (by the intercept at $l=0$).

Using the polarization angles from the R and H polarimetric datasets, we have computed 
the SF, as well as the mean measurement uncertainties $\sigma(l)$. 
Thereafter, the ADF was built to each case, as shown in Figures \ref{turbulence}a and \ref{turbulence}b, 
which are respectively related to the R and H band polarization data.
Prior to calculating the ADF, the factor $\sigma^{2}(l)$ was subtracted within each bin of the 
SF, as suggested in Equation \ref{strucfuc} (therefore, the value of each bin in Figure 
\ref{turbulence} is $\sqrt{\langle\Delta\Phi^{2}(l)\rangle - \sigma^{2}(l)}$).
In diagrams \ref{turbulence}a and \ref{turbulence}b, bin widths of respectively $2$ and $0.25$ 
arcmin have been used. 
Only the first six points of the ADF (denoted by blue open circles) are used in the 
linear fit of Equation \ref{strucfuc}, in order to guarantee the validity of the $l\ll d$ regime,
and the fitted function is represented by the red dashed line. The ratio of the turbulent to 
large-scale magnetic field strength was computed through Equation \ref{btb0}, resulting in 
$\langle B_{t}^{2}\rangle^{1/2}/B_{0} = (20.2\pm0.5)\%$ and 
$\langle B_{t}^{2}\rangle^{1/2}/B_{0} = (53\pm1)\%$ for the R and H band datasets, respectively.

\placefigure{turbulence}

The calculation shows that the turbulent component of the magnetic field is much stronger 
within the small scale area surrounding the cluster (i.e., where the H band data is distributed), 
when compared with the global turbulent contribution considering the entire area surrounding 
the H{\sc ii} region (the R band data distribution). In fact, a higher turbulence ambient
is expected to exist nearer the cluster due to the larger interaction between the young 
stars and the interstellar medium within this area. We may also associate this evidence
with our hypothesis that the low polarization values inside the H{\sc ii} region 
(mainly near the cluster, as seen in Figure \ref{mapspolthetaR}b), indeed occurs due to low polarization 
efficiency. The higher turbulence level probably represents the main factor inhibiting
the alignment of dust grains with respect to the magnetic field lines, causing the 
lower polarization values.

\section{Conclusions}

Using two sets of NIR photometric data collected using the SofI camera at the
NTT 3.5\,m telescope, we have characterized the stellar population of a deeply 
embedded stellar cluster related to the RCW41 H{\sc ii} region. Furthermore, through 
optical (V, R and I bands) and NIR polarimetric observations (H band), obtained 
respectively with similar polarimetric modules at the 0.9\,m telescope on CTIO/Chile and 
the 1.6\,m telescope on OPD/Brazil, we have built linear polarization maps of 
the star-forming area. Our main results from both techniques are listed below:

(i) Analysis of photometric colors of cluster's stars in comparison with 
field stars shows that a number of the cluster's objects present color excess, which 
is typically associated with circunstellar disks of young PMS stars. The cluster
also presents several highly reddened objects ($A_{V}>20$ mag);

(ii) A mean visual extinction of $A_{V}=7.6\pm2.0$ is estimated for the cluster's objects, 
based on the location of the majority of the stars on the color-color diagram, which 
roughly defines a band parallel to the reddened locus of CTT stars. This is corroborated 
by an independent $A_{V}$ computation using MS stars from the cluster;

(iii) The morphology of the cluster is divided between a main portion and a sub-cluster.
There seems to exist a correlation between the location of the highly obscured 
objects with the HNCO contours, which is mostly concentrated toward the north of the
cluster's main part. Therefore, these are probably highly embedded objects, and 
a local rising extinction gradient in the south-north direction is probably present
near the cluster. This evidence supports the idea that the interstellar material surrounding
Obj2 (a O9{\sc v} star at the sub-cluster, presumably the most massive object in the field) 
have already been swept out, while near the main portion, 
gas and dust are still beginning to be affected, according to the HNCO contours;

(iv) The color-magnitude diagrams of the cluster's stars show that the majority of 
the objects are displaced toward higher color values, indicating that these stars 
are mainly young objects, also presenting color excess and non-uniform interstellar
extinction. Comparison with PMS isochrones provide a mean cluster age between 
$2.5$ and $5.0$ Myr. Stars from the turn-on region (more massive) are mostly 
concentrated near the subcluster, suggesting that this area has been formed earlier, 
subsequently triggering star formation at the main portion of the cluster. Results 
from item (iii) corroborate this hypothesis;

(v) The nature of some objects previously observed through NIR spectroscopy was reviewed.
Obj1b, the close companion to the massive B0{\sc v} central object, is highly reddened and 
presents spectroscopic features similar to T Tau Sb, being therefore a very young T Tauri star.
Objects 3, 5, and 6 were previously identified as late-type field stars, although their 
position in the color-magnitude diagrams is consistent with the T Tauri classification.
Examples have been found of young objects with similar spectral features.
Specifically, Obj3 is highly reddened ($A_{V} > 20$) and very luminous, with spectral 
features that may be consistent with a medium mass YSO. Its weak CO absorption lines 
may be explained due to veiling from the circunstellar dust emission;

(vi) Using the density distribution of stars in the field, the cluster's center 
has been determined to be at
($\alpha_{\mathrm{c}}$,$\delta_{\mathrm{c}}$)$_{2000}$=($9^{\mathrm h}16^{\mathrm m}43.5^{\mathrm s}$,$-47^{\circ}56'24.2''$).
A two-parameter King function has been adjusted to the radial density profile, 
considering both the entire area around the cluster and a second fit where the 
portion encompassing the sub-cluster was removed, providing a much sharper profile. 
The general morphology shows that although some stellar scattering exist toward the south, 
the main cluster is probably of the centrally condensed-type. This shape is correlated 
to the spherical and similarly sized cloud core in which the cluster is immersed, 
as observed from the 1.2mm dust continuum emission toward RCW41.

(vii) The R-band polarimetric survey of a large area encompassing the entire H{\sc ii} region, 
was used to build images of mean polarization degree and position angle. A radial analysis 
starting from the center of the ionized area revealed a sharp increase in polarization 
degree roughly coinciding with a conspicuous interstellar ring detected 
at the MSX 8.2$\mu$m emission, which delimits the ionized area. This evidence reflects
a global decrease in polarization degree inside the H{\sc ii} region, mainly near the cluster, 
which is most probably due to low grain alignment efficiency caused by 
the turbulent environment and weak magnetic fields in this area;

(viii) The image of mean polarization angle exhibit large areas, mainly toward the southwest
of RCW41, where polarization angles are slanted outwards, following the borders of the ionized 
area. Such slanting coincides with the limits of the region indicated by the MSX 8.2$\mu$m 
emission, and is supported by simulations of the expansion of H{\sc ii} regions, which predicts 
that the magnetic field lines at the neutral gas outside the limits of the region, tends to 
lie parallel to the ionization front;

(ix) The H-band polarimetric survey, focused on a smaller area towards the cluster, 
shows that, although a large dispersion of polarization angles are present, a mean 
pattern may be noticed: polarization vectors roughly follow the direction of several 
interstellar filaments, emanating radially from the cluster, as revealed by the 
{\it Spitzer} $5.8\mu$m image. Therefore, the mean orientation is roughly perpendicular 
to the ionization front, a result which is also in agreement with the predicted 
morphology of the magnetic field lines inside the ionized area, from simulations of 
expanding H{\sc ii} regions;

(x) By combining polarimetric optical data (V, R, and I) and the NIR H-band, 
the Serkowski function has been fitted towards several stars mainly near the cluster. 
A large rotation of polarization angles is noticed toward some stars, revealing 
that the light beam has probably trespassed different interstellar layers along the 
line-of-sight. The total-to-selective extinction ratio has been computed toward each
of the fitted objects, and a mean value of $\overline{R} = 2.96\pm0.42$ has been 
obtained, which agrees with the standard $3.09$ value for the typical interstellar
medium extinction law. 

(xi) Through the statistical correlation analysis (computing the ADFs using 
the R and H band polarimetric datasets), we have shown that the ratio of the turbulent 
to large-scale magnetic field strength is $(20.2\pm0.5)\%$ when the entire area 
surrounding the H{\sc ii} region is considered, and $(53\pm1)\%$ if only the area 
near the cluster is taken into account. The higher turbulence level within the cluster 
area reinforces the hypothesis that the low polarization values at this region occurs 
due to the lower polarization efficiency.

\acknowledgements

We thank the staff of the CTIO (Chile) and OPD/LNA (Brazil) for their hospitality and 
invaluable help during our observing runs. We are also thankful to ESO for providing 
us the NIR image data used in this investigation. 

ARL thanks partial financial support from the ALMA-CONICYT Fund, under
the project number 31060004, ``A New Astronomer for the Astrophysics
Group – Universidad de La Serena", by the Department of Physics
of the Universidad de La Serena, and to the Research Director of the Universidad
de La Serena through DIULS program “Proyecto Convenio de
Desempe\~no CD11103.

This investigation made extensive use of data products from the Two Micron All Sky Survey, 
which is a joint project of the University of Massachusetts and the Infrared Processing and 
Analysis Center/California Institute of Technology, funded by the National Aeronautics and 
Space Administration and the National Science Foundation.  

We made use of Montage, funded by the National Aeronautics and Space Administration's Earth 
Science Technology Office, Computation Technologies Project, under Cooperative Agreement 
Number NCC5-626 between NASA and the California Institute of Technology. 

This research is based in part on observations made with the Spitzer Space Telescope, which 
is operated by the Jet Propulsion Laboratory, California Institute of Technology under a 
contract with NASA.

We are grateful to Drs. A. M. Magalh\~aes and A. Pereyra for providing the polarimetric unit 
and the software used for data reductions. This research has been partially supported by the
Brazilian agencies FAPEMIG and CNPq.

{\it Facilities:} 
\facility{CTIO:0.9m, LNA:1.6m}

\bibliography{astroref}
\bibliographystyle{aa}

\appendix

\section{Photometric and Polarimetric Data}
\label{apx}

The lists related to the photometric data are available in Tables 
\ref{tab_phot_A} (Dataset A) and \ref{tab_phot_B} (Datset B), and the 
complete multi-band polarimetric sample is provided in Table \ref{tab_pol}.
Tables \ref{tab_phot_A}, \ref{tab_phot_B} and \ref{tab_pol} are published in its entirety in the electronic edition of
the Astrophysical Journal. A portion of each table is shown here for guidance regarding its form and content.


\begin{table}[!h]
\scriptsize
\centering
\caption{\label{tab_phot_A} NIR phtometric data of RCW41 from Dataset A.}
\begin{tabular}{cccccccc}
\tableline\tableline
$\alpha_{2000} (^{hms})$ & $\delta_{2000} (\degr\arcmin\arcsec)$ & $J$ & $\Delta J$ & $H$ & $\Delta H$ & $K\!s$ & $\Delta K\!s$ \\ \hline
  9 16 47.94 &  -47 57 17.8 &   9.712 &  0.022  &  8.962 &  0.015  &  8.653 &  0.026 \\ 
  9 17  1.91 &  -47 56 44.7 &  10.999 &  0.020  &  9.456 &  0.018  &  8.786 &  0.016 \\ 
  9 16 43.49 &  -47 56 22.9 &  10.684 &  0.021  &  9.865 &  0.015  &  9.492 &  0.026 \\ 
  9 17  4.34 &  -47 55 49.5 &  12.801 &  0.020  & 10.852 &  0.018  &  9.959 &  0.016 \\ 
  9 16 47.67 &  -47 57 21.2 &  12.037 &  0.022  & 10.910 &  0.015  & 10.486 &  0.026 \\ 
  9 16 41.89 &  -47 56 15.4 &  14.555 &  0.038  & 12.041 &  0.015  & 10.627 &  0.026 \\ 
  9 16 47.05 &  -47 56 44.0 &  12.636 &  0.021  & 11.481 &  0.015  & 11.039 &  0.026 \\ 
  9 17  8.72 &  -47 57 16.1 &  12.982 &  0.020  & 11.764 &  0.019  & 11.183 &  0.016 \\ 
  9 16 45.57 &  -47 56 56.0 &  12.387 &  0.021  & 11.707 &  0.015  & 11.391 &  0.026 \\ 
  9 16 44.89 &  -47 55 38.6 &  13.740 &  0.021  & 12.407 &  0.015  & 11.483 &  0.026 \\ 
\hline
\end{tabular}
\tablecomments{
The columns above respectively represent each of the objects' equatorial coordinates ($\alpha,\delta$),
as well the $J$, $H$ and $K\!s$ magnitudes, together with each uncertainty ($\Delta J$, $\Delta H$ and $\Delta K\!s$).
Values marked with * represent undetected stars (in each specific band), or saturated objects.
}
\end{table}

\begin{table}[!h]
\scriptsize
\centering
\caption{\label{tab_phot_B} NIR phtometric data of RCW41 from Dataset B.}
\begin{tabular}{cccccc}
\tableline\tableline
$\alpha_{2000} (^{hms})$ & $\delta_{2000} (\degr\arcmin\arcsec)$ & $H$ & $\Delta H$ & $K\!s$ & $\Delta K\!s$ \\ \hline
  9 16 47.95 &  -47 57 17.8 &    *    &   *     &  8.621 &  0.036 \\ 
  9 17  4.88 &  -47 54  4.8 &  10.411 &  0.016  &  9.534 &  0.028 \\
  9 16 43.49 &  -47 56 22.8 &   9.944 &  0.024  &  9.543 &  0.028 \\
  9 17  4.34 &  -47 55 49.4 &  10.870 &  0.024  &  9.995 &  0.026 \\
  9 16 47.68 &  -47 57 21.2 &  10.920 &  0.016  & 10.455 &  0.021 \\
  9 16 56.95 &  -47 58 23.5 &  11.194 &  0.015  & 10.500 &  0.016 \\
  9 16 41.88 &  -47 56 15.4 &  12.068 &  0.022  & 10.666 &  0.024 \\
  9 16 25.46 &  -47 59 30.9 &  11.337 &  0.014  & 10.765 &  0.011 \\ 
  9 16 34.57 &  -47 56 58.3 &  11.650 &  0.015  & 10.814 &  0.015 \\ 
  9 16 50.51 &  -47 53 16.6 &  11.093 &  0.035  & 10.924 &  0.029 \\ 
\hline
\end{tabular}
\tablecomments{
The columns above respectively represent each of the objects' equatorial coordinates ($\alpha,\delta$),
as well the $H$ and $K\!s$ magnitudes, together with each uncertainty ($\Delta H$ and $\Delta K\!s$).
Values marked with * represent undetected stars (in each specific band), or saturated objects.
}
\end{table}

\begin{table}
\scriptsize
\centering
\caption{\label{tab_pol} The V, R, I and H polarimetric data set.}
\begin{tabular}{ccccccc}
\tableline\tableline
$\alpha_{2000} (^{hms})$ & $\delta_{2000} (\degr\arcmin\arcsec)$ & Band & $P(\%)$ & $\sigma_P(\%)$ & $\theta(^{\circ})$ & $\sigma\theta(^{\circ})$ \\ \hline
 9 15 37.95  &  -48  1 24.9  &  R  &   2.70 &  0.44 & 144.6 &   4.7 \\
 9 15 41.72  &  -47 53 10.7  &  R  &   5.33 &  1.26 &   3.3 &   6.8 \\
 9 15 43.66  &  -47 47 39.1  &  R  &  10.62 &  1.86 &   1.8 &   5.0 \\
 9 16 22.63  &  -47 57 57.1  &  V  &  10.47 &  3.34 & 171.1 &   9.1 \\
             &               &  R  &   7.97 &  1.44 & 163.9 &   5.2 \\
             &               &  I  &   6.12 &  0.99 & 162.1 &   4.6 \\
 9 16 25.00  &  -47 58  1.0  &  R  &  10.72 &  1.76 & 177.8 &   4.7 \\
             &               &  I  &   7.32 &  0.75 &   0.1 &   3.0 \\
 9 16 36.83  &  -47 56 27.8  &  H  &  12.06 &  1.59 & 111.5 &   6.3 \\
 9 16 38.41  &  -47 57 14.9  &  V  &   3.16 &  0.31 & 167.7 &   2.9 \\
             &               &  R  &   2.51 &  0.28 & 168.5 &   3.2 \\
             &               &  I  &   2.86 &  0.35 & 174.1 &   3.5 \\
             &               &  H  &   0.87 &  0.07 & 177.8 &   5.5 \\
 9 16 40.99  &  -47 55  9.9  &  V  &   3.60 &  0.61 & 138.7 &   4.9 \\
             &               &  R  &   2.76 &  0.16 & 148.0 &   1.7 \\
             &               &  I  &   2.35 &  0.09 & 150.8 &   1.1 \\
             &               &  H  &   1.03 &  0.07 & 155.4 &   5.3 \\
 9 16 42.57  &  -47 57 24.4  &  V  &   4.67 &  5.42 & 159.6 &  33.2 \\
             &               &  R  &     ** &    ** &    ** &    ** \\
             &               &  I  &   2.66 &  2.53 & 139.4 &  27.2 \\
             &               &  H  &   3.99 &  1.09 & 165.4 &   9.3 \\
 9 16 44.28  &  -47 58 17.1  &  V  &   0.85 &  0.38 & 178.0 &  12.7 \\
             &               &  R  &   0.78 &  0.37 & 173.6 &  13.7 \\
             &               &  I  &   0.44 &  0.34 & 170.7 &  22.4 \\
             &               &  H  &     ** &    ** &    ** &    ** \\
 9 16 45.47  &  -48  0  3.4  &  H  &  11.43 &  1.73 & 147.2 &   6.6 \\
\hline
\end{tabular}
\tablecomments{
The columns above respectively represent each of the objects' equatorial coordinates ($\alpha,\delta$), 
the spectral band of the polarization measurement (V, R, I, or H), the polarization degree and its
uncertainty ($P$ and $\sigma_P$), and the polarization angle with its uncertainty ($\theta$ and $\sigma\theta$).
Polarization and angle values marked with ** represent objects that were detected in the indicated band, but 
presents bad polarization values, as defined in Section \ref{observationspol}. 
}
\end{table}

\end{document}